\documentclass[aps,superscriptaddress]{revtex4}
\usepackage{amssymb,amsmath,epsfig}
\usepackage[colorlinks=true, pdfstartview=FitV, linkcolor=blue, citecolor=red, urlcolor=magenta, breaklinks=true]{hyperref}
\usepackage{graphicx}
\usepackage{epstopdf}
\usepackage[capposition=top]{floatrow}
\usepackage{subfig}

\begin{document}
\title{An Effective Model for Glueballs and Dual Superconductivity at Finite Temperature}

\author{Adamu Issifu}\email{ai@academico.ufpb.br}
\affiliation{Departamento de F\'isica, Universidade Federal da Para\'iba, 
Caixa Postal 5008, 58051-970 Jo\~ao Pessoa, Para\'iba, Brazil}

\author{Francisco A. Brito}\email{fabrito@df.ufcg.edu.br}
\affiliation{Departamento de F\'isica, Universidade Federal da Para\'iba, 
Caixa Postal 5008, 58051-970 Jo\~ao Pessoa, Para\'iba, Brazil}
\affiliation{Departamento de F\'{\i}sica, Universidade Federal de Campina Grande
Caixa Postal 10071, 58429-900 Campina Grande, Para\'{\i}ba, Brazil}

\begin{abstract}

The glueballs lead to gluon and QCD monopole condensations as by-products of color confinement. A color dielectric function $G(|\phi|)$ coupled with Abelian gauge field is properly defined to mediate the glueball interactions at confining regime after {\it spontaneous symmetry breaking} (SSB) of the gauge symmetry. The particles are expected to form through quark gluon plasma (QGP) hadronization phase where the free quarks and gluons start clamping together to form hadrons. The QCD-like vacuum $\langle\eta^2m_\eta^2F^{\mu\nu}F_{\mu\nu}\rangle$, confining potential $V_c(r)$, string tension $\sigma$, penetration depth $\lambda$, superconducting and normal monopole densities ($n_s\,n_n$) and the effective masses ($m_\eta^{2}$ and $m_A^{2}$) will be investigated at finite temperature $T$. We also calculate the strong `running' coupling $\alpha_s$ and subsequently the QCD $\beta$-function. {\it Dual superconducting} nature of the QCD vacuum will be investigated based on monopole condensation. 

\end{abstract}
\maketitle
\pretolerance10000

\section{Introduction}
The Large Hadron Collider (LHC) is currently attracting the attention of particle physicists to investigating the dynamics of the Standard Model (SM) at the TeV scale and to probe for a possible new physics. One of the major expectations at the LHC is the observation of jet substructure. Generally, jets are collimated bundles of hadrons constituted by quarks and gluons at short distances or high energies \cite{Marzani,Ali,Kogler}. Jets played a significant role in the discovery of gluons (g) \cite{JADE,PLUTO,MARK,TASSO} and top quark (t) \cite{DO,CDF}. These observations featured prominently in classifying Quantum Chromodynamics (QCD) as a theory for strong interaction within the standard model. {The most abundant secondary particles produced in heavy ion collisions are pions which emerge from hot and dense hadron gas states. These pions can be studied from the projected quark-gluon-plasma (QGP) and the deconfinement phase through emitted photons and dileptons produced at this phase. Since the photon and the dilepton do not interact with the hadron matter through strong force, they decay quickly \cite{Rapp,Li,Schulze,Volkov} making it easy to study the pions. Due to the large production of jets and pions at the LHC and Relativistic Heavy-Ion Collider (RHIC) they are suitable candidates for the search for new particles and the testing of the SM properties.} 

Quarks and gluons behave as quasi free particles at higher energies (short distances) due to asymptotically free nature of the QCD theory. When the energy among these color particles are reduced to about $1\,\text{GeV}$ and below or at a separation of order $1\,\text{fm}$ or higher, {\it color confinement} sets in, such that, the quarks and the gluons coexist as hadrons. At lower enough energies hadronization is highly favoured leading to formation of light pions. Hadron jet was first produced by annihilation of electron-positron to produce {\it two jet} event, $e^+e^-\rightarrow q\bar{q}\rightarrow 2-\text{jets}$. This {\it two jet} structure \cite{Hanson} was first observed in 1975 at SPEAR (SLAC) and the spin $1/2$ nature of the quarks also established \cite{Schwitters}. The jets are produced when the $q\bar{q}$ fly apart, more $q\bar{q}$ pairs are produced which recombine with the existing pairs to form mesons and baryons as a by-product \cite{Wu}. The first {\it three jet} structure from electron-positron annihilation process was observed by analysing $40$ hadronic events at the center-of-mass energy $27.4\,\text{GeV}$ from SPEAR \cite{Wu}. These jets were subsequently observed clearly and separately \cite{Wiik,Zobernig}. The discovery of the gluon through the  {\it three jet} process played a major role in the discovery of Higgs particle \cite{Englert} by ATLAS collaboration \cite{ATLAS} and CMS collaboration \cite{CMS} using the LHC at CERN. 
{On the other hand, pion-pion annihilation serves as the principal source of dileptons ($e^-e^+$ and $\mu^-\mu^+$) produced from the hadron matter \cite{Koch,Haglin}. Thus, the proper study of the main source of dilepton spectra observed experimentally proposes significant observables which help to understand the pion dynamics in dense nuclear matter that exist at the beginning of the collision \cite{Anchishkin}.} 

{We consider a complex scalar field model coupled to Abelian gauge fields in two different ways. 
At relatively low energies the particles undergo transformation into glueballs. {The emission of gluon $g$ which mediates the strong interactions is a non-Abelian feature observed in the nonperturbative regime of QCD theory. However, in the model framework, the gluon emission is a result of the modification of the Abelian gauge field by the color dielectric function.} 
In this approach, we approximate the non-Abelian gauge field responsible for color confinement with an Abelian gauge field coupled with a dimensionless color dielectric function $G(|\phi|)$. The color dielectric function is responsible for regulating the long distance dynamics of the photon propagator, so it does not decouple at higher wavelengths where glueballs are expected to be in a confined state. The color dielectric function is defined in terms of the glueball field $\eta$ after SSB, to give meaning to {\it color confinement} and its associated properties. While $\phi$ initiates the annihilation and production processes, $\eta$ is the glueball field which brings about confinement and gluon condensation. 
The Abelian approach to QCD theory was first proposed by 't Hooft to justify magnetic monopole condensation in QCD vacuum leading to {\it color superconductivity} \cite{Hooft1,Ezawa} that also suggested `infrared Abelian dominance'. Fast forward, it has also been shown  that about 92\% of the QCD string tension is Abelian \cite{Shiba}, so an Abelian approximation will not be out of place. Further studies of Abelian dominance for color confinement in $\text{SU}(2)$ and $\text{SU}(3)$ lattice QCD can be found in Ref.\cite{Suzuki}. An interested reader can also refer to Ref.\cite{Sakumichi} for more recent results on Abelian dominance in QCD theory. This approach has also been adopted in other confinement models ---see \cite{A-B,Adamu,Issifu} and references therein for more justifications.

The annihilation of $\phi^*\phi$ through `scalar QED' where the particles are relatively free can be used to address hadronization due to the presence of high density deconfined mater. {The spontaneous symmetry breaking of the $\text{U}(1)$ symmetry to an energy regime is suitable for describing color confinement with a glueball field $\eta$}. Because the running coupling, $\alpha_s(Q^2)$, of the strong interaction decreases with momentum increase (short distances), { $\phi^*\phi$ annihilation can be studied from perturbation theory ($\phi^*\phi\,\rightarrow\,\phi^*\phi$) similar to 
an annihilation and creation of pions ($\pi^-\pi^+\,\rightarrow\,\pi^-\pi^+\gamma$) and Bhabha scattering ($e^-e^+\rightarrow e^-e^+$ ) \cite{Louis}. 
A low energy analyses of $\phi^*\phi$ annihilation will be made to show {\it color confinement}, {\it bound states of the gluons} (glueball masses) \cite{Morningstar, Loan,Chen} and the {\it QCD vacuum} \cite{Pasechnik,Schaefer} responsible for the {\it gluon condensate}.} We will shed light on monopole condensation in the QCD vacuum and its role in screening the QCD vacuum from external electric and magnetic field penetrations similar to superconductors.} 
 
There is a long standing belief that quark masses are obtained through the spontaneous symmetry breaking mechanism facilitated by the non zero expectation value of the Higgs field. That notwithstanding, quark masses continue to be an input parameter in the standard model \cite{Melnikov}. As a free parameter in the SM, the magnitude of the mass is determined phenomenologically and the result compared with sum rules, lattice simulation results, other theoretical techniques and experimental data.
The Dyson-Schwinger integral equation for the quark mass function has two known solutions i.e., trivial and nontrivial solutions. The latter is obtained under low energy and nonperturbative conditions while the former leads to unphysical, massless quarks. Thus, the non trivial solution leads to the creation of quark masses resulting into `dynamical chiral symmetry breaking', that is a consequence of {\it confinement} \cite{Fomin,Higashijima,Acharya,Bando}. 

Monopoles can be said to be a product of grand unification theory (GUT). The mixing of the strong and the electroweak interactions due to the higher gauge symmetries in GUT breaks spontaneously at higher energies or extremely short distances. So physical features of monopoles such as size and mass are studied through the energy of the spontaneous symmetry breaking \cite{John}. Monopoles play important role in color confinement. Monopole condensation produces dual superconductor which squeezes the uniform electric field at the confinement phase into a thin flux tube picture. The flux tube is formed between quark and an antiquark to keep them confined. This scenario has been extensively investigated in lattice gauge theory \cite{Nambu} and Polyakov loop  \cite{Iwazaki} to establish its confinement properties. The involvement of monopoles in chiral symmetry has recently been investigated as well \cite{Giacomo}. The impact of monopoles in quark gluon plasma (QGP) has also been studied in \cite{Liao}.

The paper is organised as follows: In Sec.~\ref{md} we introduce the Lagrangian density that will be the basics for this study. 
In Sec.~\ref{cgmc} we present spontaneous symmetry breaking of the model presented in the previous sections; this section is divided into four subsections, in Sec.~\ref{conf} we investigate confinement of glueballs, in Sec.~\ref{em} we study the effective masses, we dedicate Sec.~\ref{gc} to gluon condensation and in Sec.~\ref{mono} we investigate the monopole condensation. We proceed to present the analysis in Sec.~\ref{ana} and the final findings in Sec.~\ref{conc}. We have adopted the natural units $c=\hbar=k_B=1$, except otherwise stated. 


\section{The Model}\label{md}
We start with a Lagrangian density developed by exploring gauge invariant properties --- see \cite{Adamu}---,
\begin{align}\label{1}
\mathcal{L}&=\eta^{\mu\nu}D_\mu\phi D_\nu\phi^*-\dfrac{1}{4}G(|\phi|)F_{\mu\nu}F^{\mu\nu}-\dfrac{1}{4}\tilde{F}_{\mu\nu}\tilde{F}^{\mu\nu}-V(|\phi|)\nonumber\\&
=\eta^{\mu\nu}(\partial_\mu\phi+iq\tilde{A}_\mu\phi)(\partial_\nu\phi^*-iq\tilde{A}_\nu\phi^*)-\dfrac{1}{4}G(|\phi|)F_{\mu\nu}F^{\mu\nu}-\dfrac{1}{4}\tilde{F}_{\mu\nu}\tilde{F}^{\mu\nu}-V(|\phi|),
\end{align}
where $D_\mu=\partial_\mu+iq\tilde{A}_\mu$, $\tilde{F}_{\mu\nu}=\partial_\mu\tilde{A}_\nu-\partial_\nu\tilde{A}_\mu$ and $F_{\mu\nu}=\partial_\mu A_\nu-\partial_\nu A_\mu$ are the covariant derivative, strength of the dual gauge field and the gauge field respectively. {This model can be used to study the annihilation and creation of identical particles $\phi^*\phi$ and its transformation into glueballs $\eta$ at low energies. The complex scalar fields $\phi(x)$ and its conjugate $\phi^*(x)$ describe particle and its antiparticle with the same mass but different charges respectively. 
 The electromagnetic interactions among the scalar fields with an exchange of single photon produced from the dual gauge field dynamics $\tilde{F}_{\mu\nu}\tilde{F}^{\mu\nu}$ is the suitable scenario to address high energy limit. However, we shall focus on the strong interactions, where the scalar fields modify to glueball fields at relatively low energies, which will lead to {\it color confinement} and its associated properties mediated by the modified gauge field dynamics $G(|\phi|)F_{\mu\nu}F^{\mu\nu}$.} 
The equations of motion are
\begin{align}\label{2}
D_\mu D^\mu\phi+\dfrac{1}{4}\dfrac{\partial G(|\phi|)}{\partial\phi^*}F_{\mu\nu}F^{\mu\nu}+\dfrac{\partial V(|\phi|)}{\partial\phi^*}=0,
\end{align}

\begin{equation}\label{2a}
\partial_\mu\tilde{F}^{\mu\nu}=-iq[\phi^*{\partial^\nu}\phi-\phi{\partial^\nu}\phi^*]+2q^2\phi^*\phi\tilde{A}^\nu=j^\nu_\phi,
\end{equation}
and 
\begin{align}\label{3}
\partial_\mu[G(|\phi|)F^{\mu\nu}]=0.
\end{align}
We will adopt a complex scalar field potential of the form,  
\begin{equation}\label{8}
V(\phi)=\dfrac{\rho}{4}[\alpha^2|\phi|^2-a^2]^2,
\end{equation}
where 
\begin{equation}\label{3aa}
\phi=\dfrac{\phi_1+i\phi_2}{\sqrt{2}}.
\end{equation}
{
We can isolate the electromagnetic interactions among the fields \cite{Dadi} for study using Eq.(\ref{2a}). This enables us to visualize how the fields annihilate and create identical particles \cite{Mandl,Peskin} through single photon exchange, $\phi(p_1)\phi^*(p_2)\rightarrow\gamma\rightarrow \phi(p'_1)\phi^*(p'_2)$, in high energy regime \cite{Langacker}.

\section{Confinement, Effective Masses, Gluon Condenstaion and Color Superconductivity}\label{cgmc}
To discuss the color confinement 
and its consequences we need to follow spontaneous symmetry breaking of the $\tilde{\text{U}}(1)$ symmetry in Lagrangian Eq.(\ref{1}) to give mass to the resulting gueball fields and its associated products such as the monopole condensation and gluon condensation which form the basics for {\it color confinement}. 
We will proceed with the symmetry breaking from the transformations,
\begin{align}\label{hm2b}
&\phi(x)\rightarrow\phi'(x)=e^{iq\alpha(x)}\phi\nonumber\\
&\tilde{A}(x)\rightarrow \tilde{A}'(x)=\tilde{A}(x)-\partial_\mu\alpha(x).
\end{align}
The vacuum expectation value of the potential Eq.(\ref{8})
\begin{equation}\label{30}
\langle |\phi| \rangle_0=\pm\dfrac{a}{\alpha},
\end{equation}
breaks the $\tilde{\text{U}}(1)$ symmetry spontaneously. 
Considering two real scalar fields $\eta(r)$ and $\zeta(r)$ representing small fluctuations about the vacuum, a shift in the vacuum can be expressed as
\begin{equation}\label{32}
\phi\rightarrow \eta(r)+\phi_0,
\end{equation} 
where $\phi_0\equiv \langle|\phi|\rangle_0$. Hence, the potential takes the form 
\begin{align}
V(\eta)&=V(\phi)|_{\langle |\phi|\rangle_0}+V'(\phi)|_{\langle |\phi|\rangle_0}\eta+\dfrac{1}{2}V''(\phi)|_{\langle |\phi|\rangle_0}\eta^2\nonumber\\
&={\rho\alpha^2a^2}\eta^2.
\end{align}
Consequently, we can suitably parametrize the scalar field such that,
\begin{align}\label{33}
 \phi&=e^{i\zeta/\phi_0}\dfrac{(\phi_0+\eta)}{\sqrt{2}}\nonumber\\
 &\approx \dfrac{(\phi_0+\eta(r)+i\zeta(r))}{\sqrt{2}}.
\end{align}
The Lagrangian in Eq.(\ref{1}) becomes, 
\begin{equation}\label{hm10}
\mathcal{L}=\dfrac{1}{2}[\partial_\mu\eta\partial^\mu\eta-2\rho\alpha^2a^2\eta^2]+\dfrac{1}{2}[\partial_\mu\zeta\partial^\mu\zeta-2q\phi_0 \tilde{A}_\mu\partial^\mu\zeta+{q^2|\phi_0|^2}\tilde{A}_\mu \tilde{A}^\mu]-\dfrac{1}{4}G(\eta)F_{\mu\nu}F^{\mu\nu}-\dfrac{1}{4}\tilde{F}_{\mu\nu}\tilde{F}^{\mu\nu}+\cdots,
\end{equation}
where $\eta$ is associated with the gueball fields and $\zeta$ is also associated with the massless Goldstone bosons which will be swallowed by the massive gauge fields through the gauge transformation \cite{Adamu}, 
\begin{equation}\label{hm12}
 \tilde{A}_\mu\rightarrow \tilde{A}'=\tilde{A}_\mu-\dfrac{1}{q\phi_0}\partial_\mu\zeta.
 \end{equation}
Pions are strongly interacting elementary particles with an integer spin. Therefore, they are bosons which are not governed by Pauli's exclusion principle and can exist as relativistic or non-relativistic particles. As a result, they can be represented by spin-0 scalar fields. In the non-relativistic regime, they exist as Goldstone bosons $\zeta$, signifying the breakdown of chiral symmetry \cite{Low}. Pions are generally produced through matter annihilations such as $p\bar{p}$, $N\bar{N}$, $ee^+$ and so on, which undergo transition from baryon structure to mesons. This is an interesting phenomenon in low energy hadron physics \cite{Dondi,Klempt}. Pions have flavour structure and other quantum numbers that permits us to classify them as bound states of quark and an antiquark. However, the valence quarks which characterizes the flavour structure are surrounded by gluons and quark-antiquark pairs. Physically, glue-rich components mix with pions ($q\bar{q}$) forming an enriched spectrum of isospin-zero states and $q\bar{q}g$ hybrid states \cite{Ochs}. The glueball spectrum and their corresponding quantum numbers are known in lattice gauge theory predictions \cite{Morningstar,Close}. Additionally, the existence of glueballs have not been decisive because of the fear of possible mixing with quark degrees of freedom \cite{Brodsky1}. In the model framework, the glueballs coexist with the Goldstone bosons which are subsequently swallowed by the massive gauge fields leaving out the glueballs ($gg$) for analysis. The glueball decay occur when the separation distance between the valence gluons exceeds some thresholds ($>1\,\text{fm}$) leading to hadronization.

Consequently, the Lagrangian can be simplified as,
\begin{equation}\label{3h}
\mathcal{L}=\dfrac{1}{2}\partial_\mu\eta\partial^\mu\eta-V(\eta)-\dfrac{1}{4}G(\eta)F_{\mu\nu}F^{\mu\nu}-\dfrac{1}{4}\tilde{F}_{\mu\nu}\tilde{F}^{\mu\nu}+\dfrac{q^2|\phi_0|^2}{2}\tilde{A}_\mu\tilde{A}^\mu,
\end{equation}
where $V(\eta)=\rho\alpha^2a^2\eta^2$. The equations of motion for this Lagrangian are
\begin{equation}\label{4a}
\partial_\mu\partial^\mu\eta+\dfrac{\partial G(\eta)}{\partial \eta}F_{\mu\nu}F^{\mu\nu}+\dfrac{\partial V(\eta)}{\partial \eta}=0,
\end{equation}
\begin{equation}\label{4b}
\partial_\mu\tilde{F}^{\mu\nu}=q^2|\phi_0|^2 \tilde{A}^\nu=j_{\phi_0}^\nu,
\end{equation}
and 
\begin{equation}\label{4c}
\partial_\mu [G(|\eta|)F^{\mu\nu}]=0.
\end{equation}
The Feynman propagator for the glueball field $\eta$ in Eq.(\ref{hm10}) reads,
\begin{equation}\label{4d1}
D_\eta(p)=\dfrac{i}{p^2-m^2_\eta+i\varepsilon},
\end{equation} 
where $m^2_\eta=2\rho\alpha^2a^2$.

\subsection{Confinement}\label{conf}

Here, we are interested in calculating static confining potential, such that only {\it chromoelectric flux} that results in confining electric field is present, whiles {\it chromomagnetic flux} influence is completely eliminated. Thus, in spherical coordinates Eq.(\ref{4a}) and Eq.(\ref{4c}) become, respectively,  
\begin{equation}\label{4d}
\dfrac{1}{r^2}\dfrac{d}{d r}\left(r^2\dfrac{d\eta}{dr}\right) =\dfrac{1}{2}\dfrac{\partial G(\eta)}{\partial\eta}E^2-\dfrac{\partial V(\eta)}{\partial\eta},
\end{equation}
and
\begin{align}\label{4e}
&\dfrac{1}{r^2}\dfrac{\partial}{\partial r}[r^2 G(\eta)E]=0 \rightarrow\nonumber\\
E&=\dfrac{\Lambda}{r^2G(\eta)},
\end{align}
where $\Lambda=q/4\pi\varepsilon_0$ is the integration constant. 
We are also now setting $G(\eta)=\xi^2V(\eta)$ {(similar process was adopted in Ref.\cite{Adamu,A-B,Issifu}), where $\xi^2$ is a dimensionful constant associated with the Regge slope $(2\pi\alpha')^2$} that absorbs the dimensionality of $V(\eta)$, so $G(\eta)$ remains dimensionless and Eq.~(\ref{4d}) reads
\begin{align}\label{5}
&\dfrac{1}{r^2}\dfrac{d}{d r}\left(r^2\dfrac{d\eta}{dr}\right)=-\dfrac{\partial}{\partial\eta}\left[\dfrac{\Lambda^2}{2}\dfrac{1}{\xi^2V(\eta)}\dfrac{1}{r^4}+V \right]\rightarrow\nonumber\\ 
&\eta''+\dfrac{2}{r}\eta'+m^2_\eta\eta=0.
\end{align}
In the last step, we have assumed that the particles are far away from the charge source $q$, 
and in such a limit terms of the order ${\cal O}(1/r^4)$ can be disregarded. 
This equation has a solution
\begin{equation}\label{12}
\eta(r)=\dfrac{a\sin(m_\eta r)}{\alpha m_\eta r}.
\end{equation}
Substituting this result into the electric field Eq.(\ref{4e}),
\begin{align}\label{13}
E&=\dfrac{\Lambda}{r^2 G}\nonumber\\
&=\dfrac{2\Lambda\alpha^2}{a^2\xi^2\sin^2(m_\eta r)},
\end{align}
and using the well known expression for determining the electrodynamic potential,
\begin{equation}\label{14}
V(r)=\int{Edr},
\end{equation}
to determine the confining potential $V_c(r)$, we get
\begin{align}\label{15}
V_c(r)&=-\dfrac{2\Lambda\alpha^2\cot(m_\eta r)}{a^2m_\eta\xi^2}+c\nonumber\\
&\simeq -\dfrac{2\Lambda\alpha^2}{a^2m_\eta\xi^2}\left[\dfrac{1}{m_\eta r}-\dfrac{m_\eta r}{3}-{\cal O}(r^3) \right]+c \nonumber\\
&\simeq \dfrac{2\alpha^2}{a^2\xi^2m^2_\eta}\left[-\dfrac{1}{r}+\dfrac{m^2_\eta r}{3} \right] ,
\end{align}
so, setting 
\begin{equation}
a^4\xi^2\rho=1\rightarrow\xi^2=\dfrac{1}{a^4\rho}
\end{equation}
leads to,
\begin{equation}\label{12a}
V_c(r)=-\dfrac{1}{r}+\dfrac{m_\eta^2 r}{3},
\end{equation}
with string tension 
\begin{equation}\label{12aa}
\sigma=\dfrac{m^2_\eta}{3}=\dfrac{2\rho\alpha^2a^2}{3}.
\end{equation}
In the last step of Eq.(\ref{15}) we substituted $\Lambda=q/4\pi\varepsilon_0=1$ and $c=0$ for simplicity.




\subsection{Effective Masses}\label{em}
Glueballs are simply bound states of gluons and they are `white' or colourless in nature. They are {\it flavour blind} and are capable of {\it decaying}. Scalar glueball with quantum number $J^{PC}=0^{++}$ is observed by lattice QCD simulations as the lightest with mass of about $1.7\,\text{GeV}$ \cite{Morningstar, Loan,Chen}.

The dispersion relation for the glueball and the gluon excitations including thermal fluctuations can be expressed as 
\begin{equation}
E_\eta^2=k^2+m^{*2}_\eta \qquad{\text{and}}\qquad E^2_A=k^2+m_A^{*2}.
\end{equation} 
We can calculate the effective thermal fluctuating glueball mass by taking the second derivative of the Lagrangian density Eq.(\ref{3h}) with respect to $\eta$,
\begin{align}\label{gb1}
m^{*2}_\eta=-\left\langle  \dfrac{\partial^2\mathcal{L}}{\partial\eta^2}\right\rangle =2\rho\alpha^2a^2+\dfrac{1}{2}\rho\alpha^2a^2\xi^2\langle F^{\mu\nu}F_{\mu\nu}\rangle.
\end{align}
We redefine the glueball field $\eta$ to include the thermal fluctuations as $\eta=\bar{\eta}+\Delta$, with restriction, $\langle\Delta\rangle=0$. The angle brackets represent thermal average and $\bar{\eta}$ is the mean field of the glueballs. This equation can be solved by determining explicitly the nature of $\langle F^{\mu\nu}F_{\mu\nu}\rangle$ and $\langle \Delta^2\rangle$ using field quanta distributions. By the standard approach, we can express,
\begin{equation}\label{gb3}
\langle F^{\mu\nu}F_{\mu\nu}\rangle=-\dfrac{\nu}{2\pi^2}\int_0^\infty{dk\dfrac{k^4}{E_A}n_B(E_A)}\quad{\text{and}}\quad \langle\Delta^2\rangle=\dfrac{1}{2\pi^2}\int_0^\infty{dk\dfrac{k^2}{E_\eta}n_B(E_\eta)},
\end{equation}
here, $n_B(x)=(e^{x\beta}-1)^{-1}$ represents the Bose-Einstein distribution function and $\beta=1/T$, where $T$ is the temperature. We can analytically solve Eq.(\ref{gb3}) at high energy regime, where $E\approx kc$, $k_BT\gg m_Ac^2$ and $m_A$ is the gluon/screening mass. 
Therefore
\begin{align}\label{gb4}
\langle F^{\mu\nu}F_{\mu\nu}\rangle&=-\dfrac{\nu}{2\pi^2}\int_0^\infty{dk \dfrac{k^4}{E_A} \dfrac{1}{e^{\beta E_A}-1} }\nonumber \\
&=-\dfrac{\nu T^4}{2\pi^2}\int_0^\infty{\dfrac{x^3dx}{e^x-1}}\nonumber\\
&=-\dfrac{4T^4}{T^4_c\xi^2},
\end{align}
where we have substituted $x=k/T$ and used the standard integral
\begin{equation}
\int{\dfrac{x^3dx}{e^x-1}}=\dfrac{\pi^4}{15}.
\end{equation}
In the last step we have substituted
\begin{equation}\label{gb5}
{T^4_c\xi^2}=\dfrac{15}{2\pi^2\nu},
\end{equation}
being $T_c$ the critical temperature. Following the same analysis, 
\begin{align}\label{gb6}
\langle\Delta^2\rangle&=\dfrac{1}{2\pi^2}\int_0^\infty{dk \dfrac{k^2}{E_\eta}\dfrac{1}{e^{\beta E_\eta}-1}}\nonumber\\
&\simeq\dfrac{T^2}{2\pi^2}\int_0^\infty{\dfrac{xdx}{e^x-1}}\nonumber\\
&\simeq\dfrac{T^2}{12},
\end{align}
where we have assumed $k_BT\gg m_\eta c^2$, with $m_\eta$ being the glueball mass, and considered the standard integral, 
\begin{equation}
\int{\dfrac{xdx}{e^x-1}}=\dfrac{\pi^2}{6}.
\end{equation}
Thus, Eq.(\ref{gb1}) becomes 
\begin{align}\label{gb7}
m_\eta^{*2}(T)&=2\rho\alpha^2a^2\left[1-\dfrac{T^4}{T^4_c} \right] \nonumber\\
&=m_\eta^2(0)\left[1-\dfrac{T^4}{T^4_c} \right].
\end{align}
When we take a thermal average of Eq.(\ref{4a}), we get $\xi^2 m_\eta^2\langle\eta F^{\mu\nu}F_{\mu\nu}\rangle=-m_\eta^2\langle\eta\rangle$, therefore, $\bar{\eta} =0$ and $\bar{\eta} =1$ are exact solutions of the glueball fields \cite{Carter}. We further demonstrate in Figs.~\ref{vacuum} and \ref{vacuum2} that $\bar{\eta}$ has solution $\bar{\eta}=0$ at $T=0$ and increases steady as $T$ increases to its maximum $\bar{\eta}\simeq 1$ at $T=T_c$. Corresponding to no glueball fields and the melting of the glueballs. 

\subsection{Gluon Condensation}\label{gc}
Classical gluodynamics is described by the Lagrangian density, 
\begin{equation}\label{gd1}
\mathcal{L}'=-\dfrac{1}{4}F^{a\mu\nu}F^a_{\mu\nu},
\end{equation}
which is invariant under scale and conformal symmetries, $x\rightarrow\lambda x$, and also produces vanishing gluon condensate $\langle F^{a\mu\nu}F^a_{\mu\nu}\rangle=0$. Meanwhile, the symmetries are broken when quantum correction $-|\varepsilon_v|$ is added to the Lagrangian i.e. 
\begin{equation}\label{gd2}
\mathcal{L}'=-\dfrac{1}{4}F^{a\mu\nu}F^a_{\mu\nu}+|\varepsilon_v|,
\end{equation}
resulting in non-vanishing gluon condensate, $\langle F^{a\mu\nu}F^a_{\mu\nu}\rangle>0$. This correction is the consequence of the well known scale anomaly observed in QCD energy-momentum ($\theta^{\mu\nu}$) trace 
\begin{equation}\label{gd2a}
\theta_\mu^\mu=\dfrac{\beta(g)}{2g}F^{a\mu\nu}F^a_{\mu\nu},
\end{equation}
where $\beta(g)$ is the `so called' beta-function of the strong coupling $g$, with a leading term
\begin{equation}\label{gd3}
\beta(g)=-\dfrac{11g^3}{(4\pi)^2}.
\end{equation}
Accordingly, we have non-zero vacuum expectation, expressed as 
\begin{equation}\label{gd4}
\langle\theta^\mu_\mu\rangle=-4|\varepsilon_v|.
\end{equation}
The second term in Eq.(\ref{3h}) acts similar to the quantum correction and explicitly breaks the scale and the conformal symmetries, so its energy-momentum tensor satisfies Eq.(\ref{gd4}) \cite{A-B,Shffman,Kharzeev,Migdal}. We will compute the trace of the energy-momentum tensor of Eq.(\ref{3h}) using the relation, 
\begin{equation}\label{gd5}
\theta_\mu^\mu=4V(\eta)+\eta\square\eta.
\end{equation}
Substituting the equation of motion Eq.(\ref{4a}) into Eq.(\ref{gd5}), yields
\begin{align}\label{gd6}
\theta_\mu^\mu&=4V(\eta)-\eta\dfrac{\partial G}{\partial\eta}F^{\mu\nu}F_{\mu\nu}-\eta\dfrac{\partial V}{\partial\eta}\nonumber\\
&=4\tilde{V}-\eta G'F^{\mu\nu}F_{\mu\nu},
\end{align}
where in the last step we have substituted $\tilde{V}=V(\eta)-\eta V'(\eta)/4$ with, $G'$ and $V'$ representing the derivative of $G$ and $V$ with respect to $\eta$ respectively. In order to make the expression easy to be compared with Eq.(\ref{gd4}) we rescale $\tilde{V}$ as $\tilde{V}(\eta)\rightarrow -|\varepsilon_v|\tilde{V}$, consequently,
\begin{equation}\label{gd7}
\langle\eta G'F^{\mu\nu}F_{\mu\nu}\rangle=4|\varepsilon_v|\langle 1-\tilde{V}\rangle.
\end{equation}
We recover the classical expression for $\langle F^{\mu\nu}F_{\mu\nu}\rangle=0$  when we set $|\varepsilon_v|\rightarrow 0$.
Following the potential defined bellow Eq.(\ref{3h}), we can express
\begin{align}\label{gd8}
\tilde{V}&=V-\dfrac{\eta V'}{4}\nonumber\\
&=\dfrac{\rho a^2\alpha^2\eta^2}{2}\nonumber\\
&=\dfrac{m^2_\eta\eta^2}{4},
\end{align}
thus, Eq.(\ref{gd7}) becomes 
\begin{equation}\label{gd9}
\langle 2G(\eta)F^{\mu\nu}F_{\mu\nu}\rangle=4|\varepsilon_v|\left\langle 1-\dfrac{m^2_\eta\eta^2}{4}\right\rangle.
\end{equation}

The QCD vacuum is seen as a very dense state of matter comprising gauge fields and quarks that interact in a haphazard manner. These characteristics are hard to be seen experimentally because quark and gluon fields can not be observed directly, only the color neutral hadrons are observable. The appearance of the mass term in the condensate is also significant because it leads to chiral symmetry break down in the vacuum. Additionally, the gluon mass $m_A$ can be derived from the vacuum  as a function of the glueball mass as 
\begin{equation}\label{gb10}
m_A^2=\dfrac{m_\eta^2}{4}.
\end{equation}
Generally, scalar glueball mass is related to the gluon mass as 
\begin{equation}
\dfrac{m(0^{++})}{m_A}=\sqrt{6}\cong 2.45,
\end{equation}
considering the leading order \cite{Cornwall} from Yang-Mills theory with an auxiliary field $\phi$, here, the scalar glueball mass $m(0^{++})$ represents fluctuations around $\phi$. 
The gluon mass is determined to be $m_A=600\sim 700\,\text{MeV}$ as obtained from lattice simulations \cite{Leinweber}. Heavier gluon masses have also been observed in the range of $\sim 1\,\text{GeV}$ by phenomenological analysis \cite{Field}, lattice simulation \cite{Leinweber1} and analytical investigations \cite{Kogan}. The glueball mass $m_\eta$ is responsible for all the confinement properties and the chiral symmetry breaking in the vacuum. Observing that the string tension is $\sigma\sim 1\,\text{GeV/fm}$, we can identify $m^2_\eta=3\,\text{GeV}^2$, corresponding to a gluon mass $m_A=0.87\,\text{GeV}$, from the model framework. 

{In terms of temperature, we use Eq.(\ref{gb6}) and the gluon condensate becomes,
\begin{align}
\langle 2G(\eta)F^{\mu\nu}F_{\mu\nu}\rangle&=4|\varepsilon_v|\left[1-\dfrac{m^2_\eta}{4}\left( \dfrac{\phi^2_0T^2}{\phi^2_0 12}\right) \right]\nonumber\\
 &=4|\varepsilon_v|\left[1-\dfrac{T^2}{T^2_{c\eta}} \right]\qquad\text{where}\qquad T_{c\eta}=\left( \dfrac{48}{m_\eta^2}\right)^{1/2}.
\end{align}
Corresponding to a temperature fluctuating gluon mass
\begin{align}
m^{*2}_A&=\dfrac{m^2_\eta T^2}{48\phi^2_0}\nonumber\\
&=\tilde{g}^2T^2,
\end{align}
where $\tilde{g}^2=m^2_\eta/48\phi^2_0$ is a dimensionless coupling constant. This expression looks like the Debye mass derived from the leading order of QCD coupling expansion \cite{Kajantie,Nadkarni,Manousakis}. It carries nonperturbative property of the theory. 
Comparing Eqs.(\ref{gd2a}) and (\ref{gd6}), we identify
\begin{align}
\dfrac{\beta(g)}{2g}=-\eta G'(\eta)\rightarrow\nonumber\\
\beta(1/r^2)=-2g\eta G'(\eta).
\end{align}
Additionally, we can determine the strong running coupling $\alpha_s$ from the renormalization group theory \cite{Deur}
\begin{align}\label{rn1}
\beta(Q^2)=Q^2\dfrac{d\alpha_s}{dQ^2},
\end{align}
therefore,
\begin{equation}
\beta(\eta)\simeq-\eta\dfrac{d(G)}{d\eta}=-\xi^2m_\eta^2\eta^2(r)=2G(\eta)=\beta(1/r^2) \qquad\text{with}\qquad g=1.
\end{equation}
Comparatively, the strong running coupling can be identified as $\alpha_s(\eta)=G(\eta)=\alpha_s(1/r^2)$, with $Q\equiv 1/r$, as the space-like momentum associated with the four vector momentum as $Q^2\equiv-q^2$. Thus, the color dielectric function is associated with the QCD $\beta$-function and the strong running coupling. 

}
We can write the Feynman propagator for the interaction by considering the left hand side of Eq.(\ref{gd7}),
\begin{align}\label{gd11}
F^{\mu\nu}F_{\mu\nu}&=-2A^\nu[\square-\partial_\mu\partial_\nu]A^\mu\nonumber\\
&=-2A^\nu\left[ \partial^2g^{\mu\nu}-\left(1-\dfrac{1}{\alpha} \right) \partial_\mu\partial_\nu\right]A^\mu \nonumber\\
&=-2A_\nu(k)\left[k^2g^{\mu\nu}-\left(1-\dfrac{1}{\alpha} \right) k^\mu k^\nu \right]A_\mu(-k) \rightarrow\nonumber\\
\eta G'(\eta)F^{\mu\nu}F_{\mu\nu}&=-2A_\nu(k)\eta G'(\eta)\left[k^2g^{\mu\nu}-\left(1-\dfrac{1}{\alpha} \right) k^\mu k^\nu \right]A_\mu(-k)\nonumber\\
&=A_\nu(k)\beta(q^2)\left[k^2g^{\mu\nu}-\left(1-\dfrac{1}{\alpha} \right) k^\mu k^\nu \right]A_\mu(-k),
\end{align}
where we have added the gauge fixing term $(2/\alpha)(\partial_\mu A^\mu)^2$, and made the Fourier transform 
\begin{equation}
A_\mu(k)=\int{\dfrac{d^4k}{(2\pi)^4}e^{-ikx}A_\mu(x)}.
\end{equation}
In the last step, the term in the square brackets is precisely the photon propagator normally associated with Abelian gauge fields but the multiplicative factor $-2\eta G'(\eta)=\beta(1/r^2)\equiv\beta(q^2)$ is defined such that the photon propagator does not decouple at higher wavelengths. 
This influences the photons to behave like gluons. 


{Also, one can retrieve all the major results obtained in \cite{A-B}, if we consider temperature fluctuations in the scalar field $\phi$. In that case, we move away from the known form of field theory i.e. perturbation about the vacuum, $\langle\phi\rangle=0$, to account for the density of particles and their interactions about the vacuum and the thermal distributions in the scalar field. So, we can substitute $\phi\rightarrow \phi+\phi_T$ into the potential in Eq.(\ref{8}), where $\phi_T$ plays a similar role as $\Delta$ introduced above. Thus, 
\begin{equation}
\langle\phi^2_T\rangle=\dfrac{1}{2\pi^2}\int_0^{\infty}{\dfrac{p^2dp}{E(e^{E\beta}-1)}}\simeq \dfrac{T^2}{12},
\end{equation}
here, we have used the high temperature approximation $k_BT\gg m_\phi c^2$, $E\approx pc$ and the potential becomes,
\begin{align}\label{sf1}
V(\phi,T)&=\dfrac{\rho}{4}[\alpha^2|\phi|^2+\alpha^2\langle\phi^2_T\rangle-a^2]^2\nonumber\\
&=\dfrac{\rho}{4}\left[\alpha^2|\phi|^2-a^2\left( 1-\dfrac{T^2}{T^2_{c\phi}}\right)  \right]^2\nonumber\\
&=\dfrac{\rho}{4} [\alpha^2|\phi|^2-\tilde{a}^2]^2,
\end{align}
where $\langle\phi_T\rangle=0$, 
\begin{equation}
T^2_{c\phi}=\dfrac{12a^2}{\alpha^2} \qquad{\text{and}}\qquad \tilde{a}^2=a^2\left(1-\dfrac{T^2}{T^2_{c\phi}} \right). 
\end{equation}
The glueball mass can be calculated from the fluctuations in Eq.(\ref{sf1}) around the vacuum, such that, 
\begin{align}\label{sf2}
m_\phi^{*2}(T)&=\dfrac{\partial^2V}{\partial\phi^2}|_{\phi_0}=2\rho\alpha^2\tilde{a}^2\nonumber\\
&=2\rho\alpha^2a^2\left[1-\dfrac{T^2}{T^2_{c\phi}} \right] \nonumber\\
&=m^2_\eta(0)\left[1-\dfrac{T^2}{T^2_{c\phi}} \right],
\end{align}
where $\phi_0=\tilde{a}/\alpha$ is the vacuum of the potential. We observe that $m_\phi^{*2}(0)=m^{*2}_\eta(0)$ at $T=0$, similar to Eq.(\ref{gb7}). The differences in the degree of the temperature in both equations arises because Eq.(\ref{gb7}) has temperature correction to the gauge field $\langle F^{\mu\nu}F_{\mu\nu}\rangle$, while in Eq.(\ref{sf2}) we have temperature correction to the scalar field $\langle\phi^2_T\rangle$. Following the same procedure as the one used in deriving Eq.(\ref{gd9}) we can express
\begin{align}\label{sf3}
\langle 2G(\eta)F^{\mu\nu}F_{\mu\nu}\rangle&=-4|\varepsilon|\left\langle\dfrac{m^{*2}_\phi(T)\eta^2}{4}-1 \right\rangle\nonumber\\
 &=-4|\varepsilon|\left\langle\dfrac{m_\phi^2(0)}{4}\left[1-\dfrac{T^2}{T^2_{c\phi}} \right]\eta^2-1 \right\rangle.
\end{align}
Here, we have replaced $m_\eta^{*2}(T)$ with $m_\phi^{*2}(T)$ from Eq.(\ref{gd9}). The negative sign and $\eta^2$ differences between this result and Ref.\cite{A-B} might be due to the differences in the methods adopted for calculating the energy-momentum trace tensor $\theta_\mu^\mu$. Also, if we consider the IR part of the potential in Eq.(\ref{15}) as studied in Ref.\cite{A-B}, the potential becomes 
\begin{align}
V_c(r)&=\sigma r=\dfrac{m_\phi^2}{3}r\rightarrow\nonumber\\
V_c(T,r)&=\dfrac{m_\phi^{*2}(T)}{3}r=\dfrac{m_\phi^2(0)}{3}\left[1-\dfrac{T^2}{T^2_{c\phi}} \right]r,
\end{align}
with string tension 
\begin{equation}\label{stringt}
\sigma(T)=\dfrac{m_\phi^{*2}(T)}{3},
\end{equation}
which is similar to the result in Ref.\cite{A-B}. Therefore, the approach adopted in this paper accounts for the temperature corrections to both the glueball field $\langle\Delta^2\rangle$ in Eq.(\ref{gb6}) and the gauge field $\langle F^{\mu\nu}F_{\mu\nu}\rangle$, while in Ref.\cite{A-B} only the thermal correction to the scalar field $\langle\phi_T^2\rangle$ was considered.

} 

\section{Dual Superconductivity}\label{mono}
{In this section we will discuss briefly {\it color superconductivity} and proceed to elaborate on {\it dual superconductivity} in detail. The fields $\phi^*\phi$ annihilate and create identical fields through a decay process, $\phi(p_1)\phi^*(p_2)\rightarrow\phi(p'_1)\phi^*(p'_2)$. The dual gauge field $\tilde{F}^{\mu\nu}\tilde{F}_{\mu\nu}$ which mediates the interaction of the scalar fields $\phi$ during the annihilation process is also responsible for the monopole condensation. In the model framework, the scalar field can be seen as point-like, asymptotically free and degenerate at the high energy regime \cite{Lan}. {In the context of this work, high energy regime is in reference to the phase at which the scalar fields are annihilating while low energy regime is after the symmetry breaking or the glueball regime.} In effect, multi-particle states are formed in the high energy regime. In such a high particle density regime, the charged scalar fields form Bose-Einstein condensation \cite{Liu1,Adhikari} similar to induced isospin imbalance systems leading to color superconductivity \cite{He,Alford1}. With this type of condensation, there is higher occupation number at the ground state than the excited states. So temperature increase goes into reducing the occupation numbers \cite{Brito}. There are several works on nonzero isospin chemical potential ($\mu_I$) and baryon chemical potential ($\mu_B$) in pion condensation \cite{Barducci2,Nishida,Hands} and they are related to the two flavor quarks that constitute the pions as $\mu_I=(\mu_u-\mu_d)/2$ and $\mu_B=(\mu_u+\mu_d)/2$ respectively. It has also been established \cite{He1} analytically at the quark level that the critical isospin chemical potential \cite{Toublan,Barducci} for pion superfluidity is precisely the pion mass $\mu_I^c=m_\pi$ in vacuum. This behaviour and other related isospin chemical potential in pion condensate have been investigated using  Nambu-Jona-Lasinio (NJL) model, ladder-QCD \cite{Barducci2}.

In the model framework, the color superconducting property can be studied from the propagator, $\langle\phi\phi^*\rangle$. High density region where the particles are asymptotically free, forming deconfined matter leading to Bose-Einstein condensation resulting into {\it color superconducting phase} \cite{Sadzikowski,Alford1,Rapp1}. On the other hand, at low energy region, the fields modify into glueballs with real mass, $m_\eta$, leading to chiral symmetry breaking and color confinement. 
Since quark-quark combinations do not form color singlet, Cooper pair condensate will break the local color symmetry giving rise to {\it color superconductivity} \cite{Alford}. 
This phenomenon was first studied by Bardeen, Cooper, and Schrieffer (BCS) \cite{Cooper}. The BCS mechanism for pairing in general, appears to be more robust in dense quark mater than superconducting metals due to the nature of interactions among quarks. The virtual fermions in the vacuum makes it {\it color diamagnetic} preventing both the electric and magnetic fields from penetrating \cite{Ali}.

We turn attention to the {\it dual superconductivity} in color confinement whose property we can derive from Eq.(\ref{4b}). This subject was first proposed by Nambu, 't Hooft and Mandelstam \cite{Nambu}. While the {\it color superconducting phase} takes place at the deconfined matter regime, the {\it dual superconductivity} is observed at the confining phase \cite{Carmona}. In the model framework, {\it dual superconductivity} is observed at high glueball condensation region, $\eta\rightarrow 0$, i.e. a region of strong confinement.  We will use the rest of this section discussing {\it dual superconductivity} in detail within the model framework. In this phenomena the QCD vacuum is seen as a color magnetic monopole condensate resulting into one dimensional squeezing of the uniform electric flux that form on the surface of the quark and antiquark pairs by {\it dual Meissner effect}. This results into the formation color flux tube between quark and an antiquark used in describing string picture of hadrons \cite{Sakumichi}. In this regard, magnetic monopole condensation is crucial to color confinement and {\it dual superconductivity}. Monopoles also play a role in chiral symmetry breaking \cite{Ramamurti} and strongly coupled QGP. They are involved in the decay of plasma formed immediately after relativistic heavy ion collision \cite{Liao}. 

Even though monopoles have not been seen or proven experimentally, there are many theoretical bases for which one can belief their existence. The theoretical evidence for its existence is as strong as any undiscovered theoretical particle. Polyakov \cite{Polyakov} and 't Hooft \cite{Hooft} discovered that monopoles are the consequences of the general idea of unification of fundamental interactions at short distances (high energies). 
Dirac on the other hand demonstrated the presence of monopoles in QED. Monopoles in general play an insightful role in understanding {\it color confinement}. They also provide some useful explanations to the features of {\it superstring theory} and {\it supersymmetric quantum theory} where the use of duality is a common phenomenon. Discovery of these monopoles some day, will be interesting and possibly revolutionary since almost all existing technologies are based on electricity and magnetism.

Monopole condensation can now be exploited from the condensation of particles in the region $r\rightarrow r_*$, where $\langle\phi\phi^*\rangle\sim \langle|\phi|^2\rangle_0$. Hence, from Eq.(\ref{4b}) we can express, 
\begin{equation}\label{mc1}
\partial_\mu\tilde{F}^{\mu\nu}=q^2|\phi_0|^2\tilde{A}^\nu=j^\nu_m.
\end{equation}
The equations of motion for the static fields are,
\begin{equation}\label{mc2}
\nabla\cdot\vec{\tilde{B}}=\rho_m \qquad{\text{and}}\qquad -\nabla\times\vec{\tilde{E}}=\vec{j}_m,
\end{equation}
here, $\vec{\tilde{E}}$ and $\vec{\tilde{B}}$ are the dual versions of the static electric and magnetic fields, while $\vec{j}_m $ and $\rho_m$ being the magnetic current and charge densities respectively. The Lorentz force associated with these fields can be expressed as
\begin{equation}\label{mc2a}
\vec{F}=q(\vec{\tilde{B}}-\vec{v}\times\vec{\tilde{E}}),
\end{equation}
where $\vec{v}$ is the speed of the particle within the fields and $q$ is the monopole charge. The homogeneous equation $\nabla\cdot \vec{\tilde{E}}=0$, represents the uniform electric field present at the confining phase. Combining the dual version of the Ampere's law on the right side of Eq.(\ref{mc2}) together with the dual London equation responsible for persistent current generated by the monopoles
\begin{equation}\label{mc3}
\nabla\times\vec{j}_m=\dfrac{1}{{\lambda}^2}\vec{\tilde{E}},
\end{equation}
we obtain the expression 
\begin{equation}\label{mc3a}
\nabla^2\vec{\tilde{E}}=\dfrac{1}{{\lambda}^2}\vec{\tilde{E}}.
\end{equation}
Noting that $\tilde{{\bf E}}\,=\,\nabla\times{\bf A}$, we can identify, ${\lambda}=(q^2|\phi_0|^2)^{-1/2}=(q^2m^2_\eta/2\alpha^4\rho)^{-1/2}$ as the London penetration depth \cite{Singh,Bazeia}. Developing the Laplacian in Eq.(\ref{mc3a}) in spherical coordinates yields,
\begin{equation}\label{mc3b}
\dfrac{d^2{\tilde{E}}_r}{dr^2}+\dfrac{2}{r}\dfrac{d\tilde{E}_r}{dr}-\dfrac{\tilde{E}_r}{{\lambda}^2}=0,
\end{equation}
this equation has a solution given as 
\begin{equation}\label{mc3c}
\tilde{E}_r=\dfrac{c_1e^{-r/{\lambda}}}{r},
\end{equation}
where $c_1$ is a constant. Using dimensional analysis, we can fix the constant as $c_1=\tilde{E}_0$, therefore, 
\begin{align}\label{mc3d}
{\tilde{E}}_r&=\dfrac{\tilde{E}_0 e^{-r/{\lambda}}}{r},
\end{align}
where ${\tilde{E}}=|{\bf\tilde{E}}|=\tilde{E}_r$. This means the electric field is exponentially screened from the interior of the vacuum with penetration depth ${\lambda}$, a phenomenon known as the {\it color Meissner effect}. Also, using London's accelerating current relation, 
\begin{equation}\label{mc3e}
\vec{\tilde{B}}=-{\lambda}^2\dfrac{\partial \vec{j}_m}{\partial t},
\end{equation}
and writing the continuity equation in the form of London equations,
\begin{equation}
\dfrac{\partial \vec{j}_m}{\partial t}+\nabla\rho_m=0,
\end{equation}
together with the equation at the left side of  Eq.(\ref{mc2}), we obtain,
\begin{equation}\label{mc3g}
\nabla^2\vec{\tilde{B}}=\dfrac{\vec{\tilde{B}}}{{\lambda}^2}.
\end{equation}
This equation has a solution similar to Eq.(\ref{mc3b}) i.e.,
\begin{equation}\label{mc3h}
\tilde{B}_r= \dfrac{\tilde{B}_0 e^{- r/\lambda}}{r}.
\end{equation}
Thus, the magnetic field is equally screened exponentially from the interior of the vacuum by a penetration depth $\lambda$. In sum, both time varying electric and magnetic fields are screened with the same depth $\lambda$. 

Now, comparing ${\lambda}$ with the results in Refs.\cite{Singh,Tinkham}, where 
\begin{equation}\label{mc4}
{\lambda}^2=\dfrac{m}{n_se^2},
\end{equation}
we rearrange the expression obtained for $\lambda$ in the model framework,
\begin{equation}\label{mc31}
\lambda^2=\dfrac{2\alpha^4\rho}{q^2m_\eta^2}=\dfrac{m^2_\eta}{2q^2\rho a^4},
\end{equation}
to make it suitable for comparison. Juxtaposing these two equations leads to, $q\,=\,e$, $m\,=\,m_\eta^2 /a$ and $n_s\,=\,2\rho a^3$, where $m,\,e,\,\text{and}\,n_s$ are mass, charge and electron number density of a superconducting material. Hence, the electron charge $e$ is equivalent to the monopole charge $q$ \cite{Dirac,Schwinger} and the electron number density and mass are related to the monopole density and mass respectively. 
Combining Eq.(\ref{mc3}) with the electric field expression at the right side of Eq.(\ref{mc2}) we obtain a {\it fluxoid quantization} relation
\begin{equation}\label{mc6}
\int{\vec{\tilde{E}}\cdot d\vec{S}}-{\lambda}^2\oint{\vec{j}_m\cdot d\vec{l}}=n\Phi_e,
\end{equation}
where $n\in\mathbb{Z}$ and $\Phi_e$ is the quantum of the electric flux \cite{Singh}. The penetration depth could also be expressed as a function of temperature,
\begin{align}\label{mc6a}
\lambda(T)&\simeq\left( \dfrac{2\alpha^4\rho}{q^2m_\eta^{*2}(T)}\right)^{1/2}\nonumber\\
&\simeq \lambda(0)\left[1-\dfrac{T^4}{T_c^4} \right]^{-1/2}. 
\end{align}
We retrieve the initial penetration depth below Eq.(\ref{mc3a}), if we set $T=0$. Accordingly, the penetration depth increases with temperature until it becomes infinite at $T=T_c$, where deconfinement and restoration of chiral symmetry coincides \cite{Gupta,Gottlieb}. At this point the field lines are expected to penetrate the vacuum causing it to loose its superconducting properties. This expression is precisely the same as the one obtained for metallic superconductors \cite{Tinkham}. At $T\,>\,T_c$ the magnetic and electric field lines goes trough the vacuum. This phenomenon is referred to as the {\it Meissner effect}. 

{The number density of the superconducting monopoles can also be expressed in terms of temperature as 
\begin{align}
n_s(T)&=2\rho a^3\left[1-\dfrac{T^4}{T_c^4} \right]^{3/2} \nonumber\\
&=n\left[1-\dfrac{T^4}{T_c^4} \right]^{3/2}.
\end{align}
The number density decreases sharply with temperature $T$ and vanishes at $T=T_c$. On the other hand, at $T<T_c$, similar to the two fluid model where normal monopole fluid mix with superconducting ones, the density of the monopoles can be calculated from the relation $n=n_s+n_n$, where $n_n$ is the normal monopole density \cite{Poole}. Additionally, at $T\rightarrow 0$, $n_s\rightarrow n$, and in the limit $T\rightarrow T_c$, $n_s\rightarrow 0$. Consequently, 
\begin{equation}
n_n=n\left\lbrace  1- \left[1-\dfrac{T^4}{T_c^4} \right]^{3/2}\right\rbrace.
\end{equation}
Thus, the {\it color superconducting} monopoles $n_s$ at $T\,<\,T_c$, conduct with dissipationless flow, on the other hand, the normal monopoles $n_n$ conduct with finite resistance at the same temperature range. Consequently, $n_s$ decreases sharply with increasing $T$ while $n_n$ increases steadily with increasing $T$.
}

\section{Analysis and Conclusion}
\subsection{Analysis}\label{ana}
{At high energy regime, there is high particle density which form deconfined matter leading to hadronization \cite{Letessier,Senger}, $\phi^*(p_1)\phi(p_2)\rightarrow\gamma\rightarrow \phi^*(p'_1)\phi(p'_2)$ \cite{Langacker,Linnyk,Liu,Bratkovskaya,Srinivasan}. 
In this regime, the process takes place at low temperature but in high momentum and particle density. This leads to an asymptotic freedom behaviour among the particles because color confinement depends on how far or close the particles are together. Thus, higher density means the particles are closer together leading to deconfinement. Additionally, Eq.~(\ref{gd11}) represents the gluon propagator which mediates the interaction of the glueballs in the IR regime of the model. Hadronization \cite{Webber} comes into play in this regime when the momenta of the particles are reduced below a particular threshold set by the QCD scale, $\Lambda_{QCD}\sim 250\,\text{MeV}$, or the separation distance between particle and antiparticle pairs is grater than $\sim 1\,\text{fm}$. 
For the sake of clarity, there are two forms of hadronization mentioned here, one arising from high energy QGP formation and the other from QCD string decay into hadrons in the low energy regime. The QGP hadronization is believed to have been formed immediately after the Big Bang when the QGP starts cooling down to Hagedorn temperature, $T\sim 170\,\text{MeV}$, or higher baryon densities above $5\rho_0$ where the free quarks and gluons start clamping together into hadrons. This is also observed at the initial stages of heavy-ion-collisions \cite{Senger,Braun1,Wagner}. On the other hand, the string decay hadronization occur due to nonperturbative effects. 
Under this picture, new hadrons are formed out of quark-antiquark pairs or through gluon cascade \cite{Webber}.}

Theoretically, Light-Front (LF) is one of the suitable theories for describing relativistic interactions due to its natural association with the light-cone. Hence, under this theory one can perform Fock expansion
\begin{equation}
|\pi\rangle=|q\bar{q}\rangle+|q\bar{q}\,q\bar{q}\rangle+|q\bar{q}\,g\rangle+ \cdots,
\end{equation}
where the valence $|q\bar{q}\rangle$ and the non valence quark components are fully covered \cite{Brodsky,Salme}. Hence, there is many information on the partonic structure of hadrons stored in the pions produced during hadronization that can be further exploited. These processes give an insight into the transition to the hypothetical QGP phase of matter where deconfinement and chiral symmetry restoration are expected. These studies gained more attention when ultra-relativistic heavy ion collision experiment in BNL Brookhaven and CERN Geneva were able to create matter under extreme conditions of temperature and densities necessary for phase transition \cite{Braun}. Pion-pion annihilation is particularly necessary since pions are the secondary most abundant particles produced during the heavy ion collision from hot and dense hadron gas. Such annihilation also result in the production of photon and dilepton which form the basics for probing quark-gluon plasma phase and further hadronizations. These particles are known to leave the dense matter phase quickly without undergoing strong interactions \cite{Rapp,Li,Schulze}. However, the leptons are capable of annihilating into quarks and gluons (partons) in a process popularly referred to as {\it gluon bremsstralhaung} \cite{Wiik,Wu}. 
 
The free parameters of the model are $\nu$, $\xi$, $|\varepsilon_v|$, $\rho$, $\alpha$, $a$ and $\phi_0$, where $\nu$ is the degeneracy of the gluons and its acceptable value for massless gluons in $\text{SU}(3)$ representation is $\nu=16$ while its value under $\text{SU}(2)$ representation is $\nu=6$. It is known from QCD lattice simulation that $\sigma\sim 1\,\text{GeV/fm}$, therefore, we can determine from Eq.(\ref{12aa}) that $m_\eta=1.73\,\text{GeV}$ \cite{Morningstar,Loan,Chen,Lee,Bali} which is precisely the scalar glueball mass. 
For zero degeneracy i.e. $\nu=0$, the critical temperature Eq.(\ref{gb5}) goes to infinity and all the glueballs get melted leaving pure gluons \cite{Carter}. We determined the gluon mass to be $m_A=870\,\text{MeV}$ in the model framework. This value lies within the values obtained from lattice simulation as referred to in Sec.~\ref{gc}. Also $|\varepsilon_v|$ is the magnitude of the QCD vacuum at the ground state. It has different values depending on the model under consideration. Its value estimated from sum rule for gluon condensate is $0.006\pm 0.012\,\text{GeV}^4$ \cite{Ioffe} and the Bag model is also quoted as $(145)^4\,\text{MeV}^4$. From the model $|\varepsilon_v|=\langle F^{\mu\nu}F_{\mu\nu}\rangle=\xi^{-2}=\rho a^4$, recalling that $\phi_0$ is related to $a$ and $\alpha$ as $\phi_0=a/\alpha$, 
$a$ has the dimension of energy and $\alpha$ is dimensionless. 

Again, when we combine Eqs.(\ref{12aa}) and (\ref{gb7}) we can express the string tension $\sigma$ as 
a function of temperature $T$
\begin{align}\label{ana1}
\sigma(T)=\dfrac{m^{*2}_\eta(T)}{3}=\dfrac{m_\eta^2(0)}{3}\left[1-\dfrac{T^4}{T^4_c} \right].
\end{align}
Here, we retrieve Eq.(\ref{12aa}) at $T=0$, $\sigma(T=T_c)=0$, and at $T\,>\,T_c$ we move into quark-gluon-plasma (QGP) phase where the particles interact in a disorderly manner \cite{Yagi,Pasechnik}. Thus, the potential in Eq.(\ref{12a}) can be expressed in terms of temperature as
\begin{align}\label{ana2}
V_c(r,T)&=-\dfrac{1}{r}+\dfrac{m_\eta^{*2}(T)}{3}r\nonumber\\
&=-\dfrac{1}{r}+\dfrac{m_\eta^2(0)}{3}\left[1-\dfrac{T^4}{T^4_c}\right] r,
\end{align}
where we recover Eq.(\ref{12a}) at $T=0$. Thermal deconfinement and the restoration of chiral symmetry occur at $T=T_c$ \cite{Gupta,Gottlieb} due to the dissolution of the glueball mass, and we have QGP phase at $T\,>\,T_c$ \cite{Pasechnik,Yagi} where the glueball mass becomes unstable. 
As established in $\text{SU}(2)$ group representation, it is known from hadron spectrum that the $\text{SU}(2)_L\times \text{SU}(2)_R$ chiral symmetry is spontaneously broken due to the nonperturbative dynamics of the theory, leading to color confinement. 
Considering quarks representation in $\text{SU}(2)$, the color charges may be screened by the gauge degrees of freedom. Under this representation, we can possibly have quark-gluon color singlet thermionic states, quark-antiquark color singlets or gluon-gluon color singlets. Generally, in this paper, we investigate glueballs which will fall under the gluon-gluon color singlet states. 

\begin{figure}[H]
  \centering
 {\includegraphics[scale=0.6]{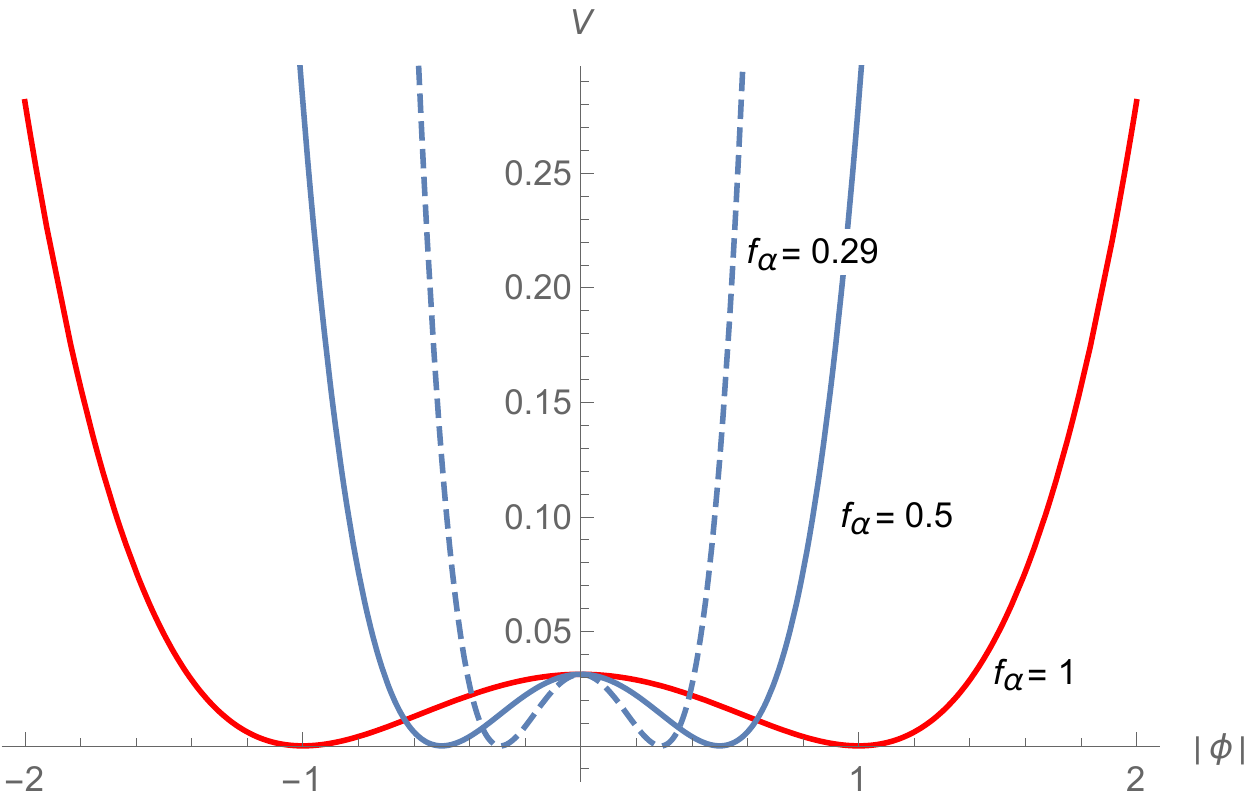}}
 \caption{The scalar field potential for different decay constants $f_\alpha$}
   \label{sp}
   \floatfoot{The potential has degenerate vacua. 
   }
\end{figure}

\begin{figure}[H]
 \centering
 {\includegraphics[scale=0.6]{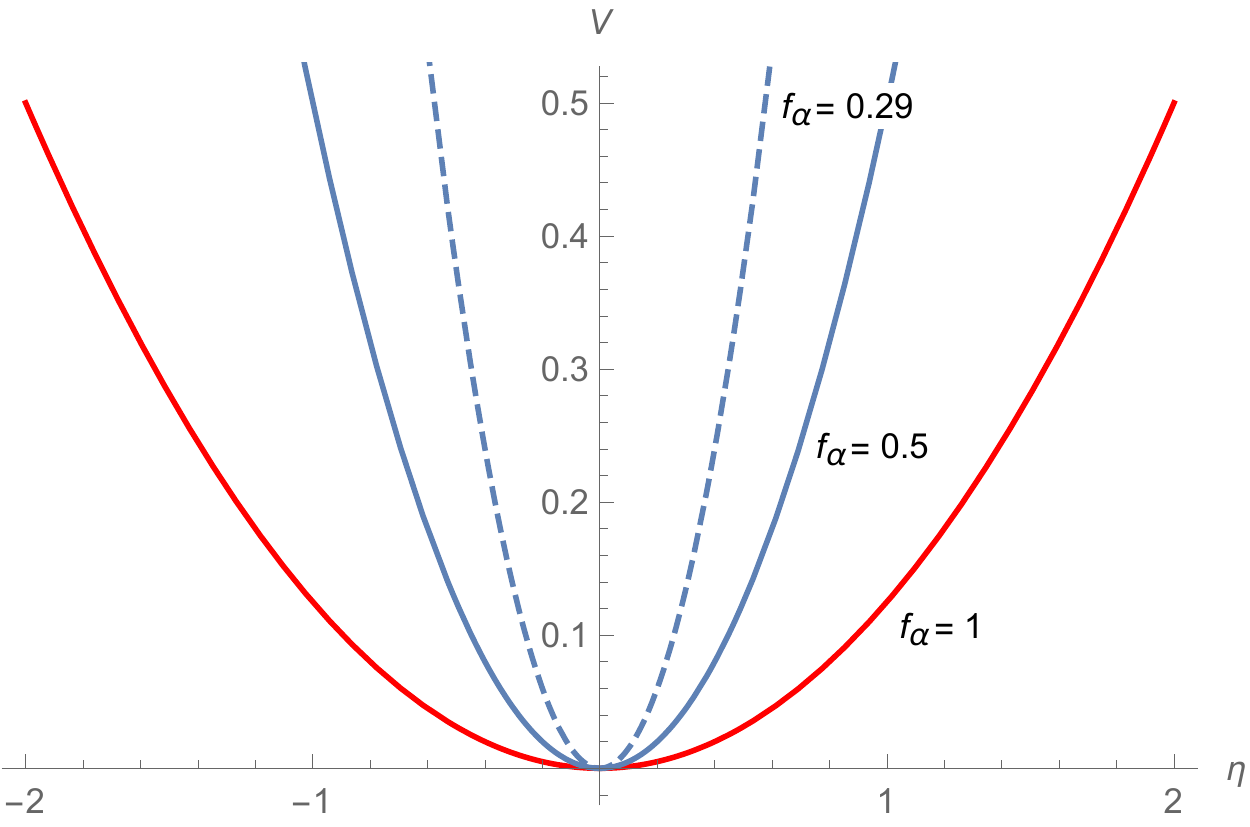}}
 \caption{The glueball potential $V(\eta)$ against the glueball field $\eta$ for different values of $f_\alpha$}
   \label{ep}
   \floatfoot{The potential has a non degenerate and well defined vacuum 
   giving the glueballs a real mass $m_\eta$. 
   }
\end{figure}

\begin{figure}[H]
  \centering
 {\includegraphics[scale=0.6]{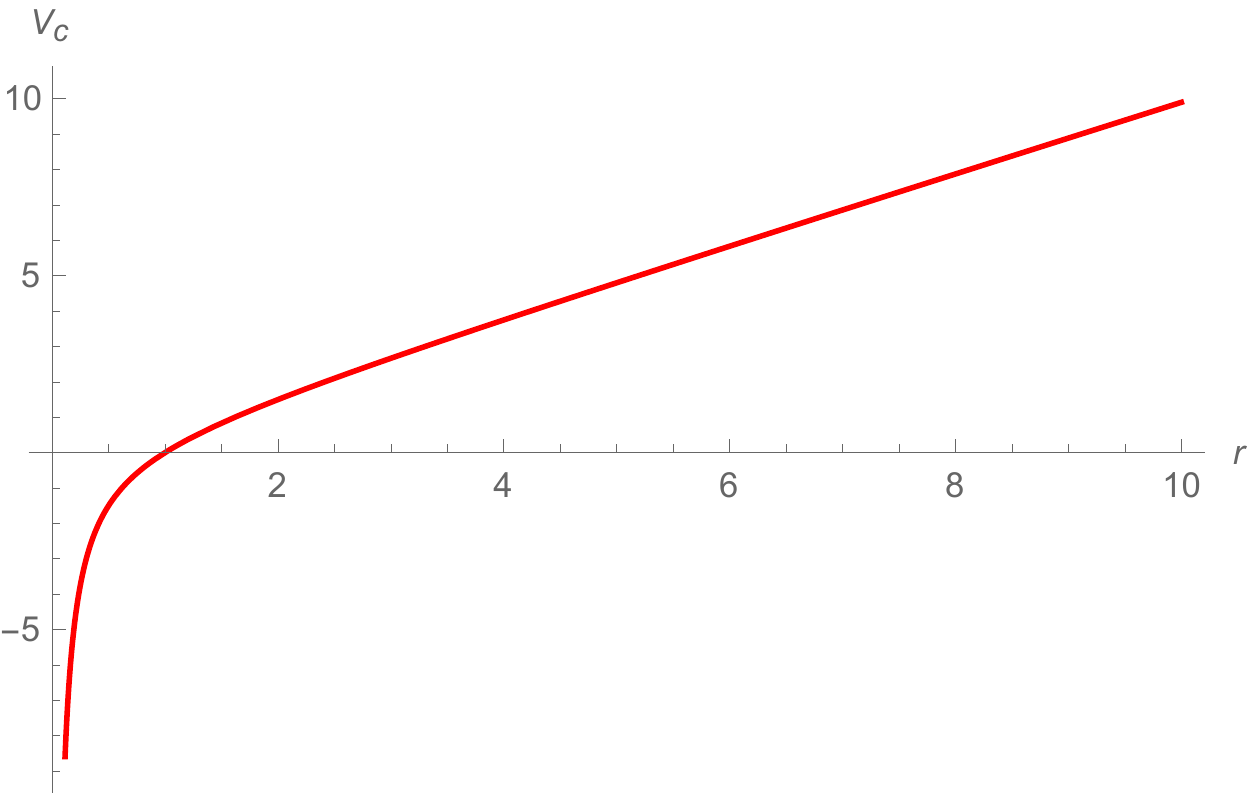}}
 \caption{Confining Cornell-like potential}
   \label{cp}
   \floatfoot{The perturbative and the nonperturbative nature of the potential is displayed. Bellow the Fermi scale the gluons are asymptotically free while above the Fermi scale we only find confined color neutral particles.}
\end{figure}

\begin{figure}[H]
  \centering
 {\includegraphics[scale=0.6]{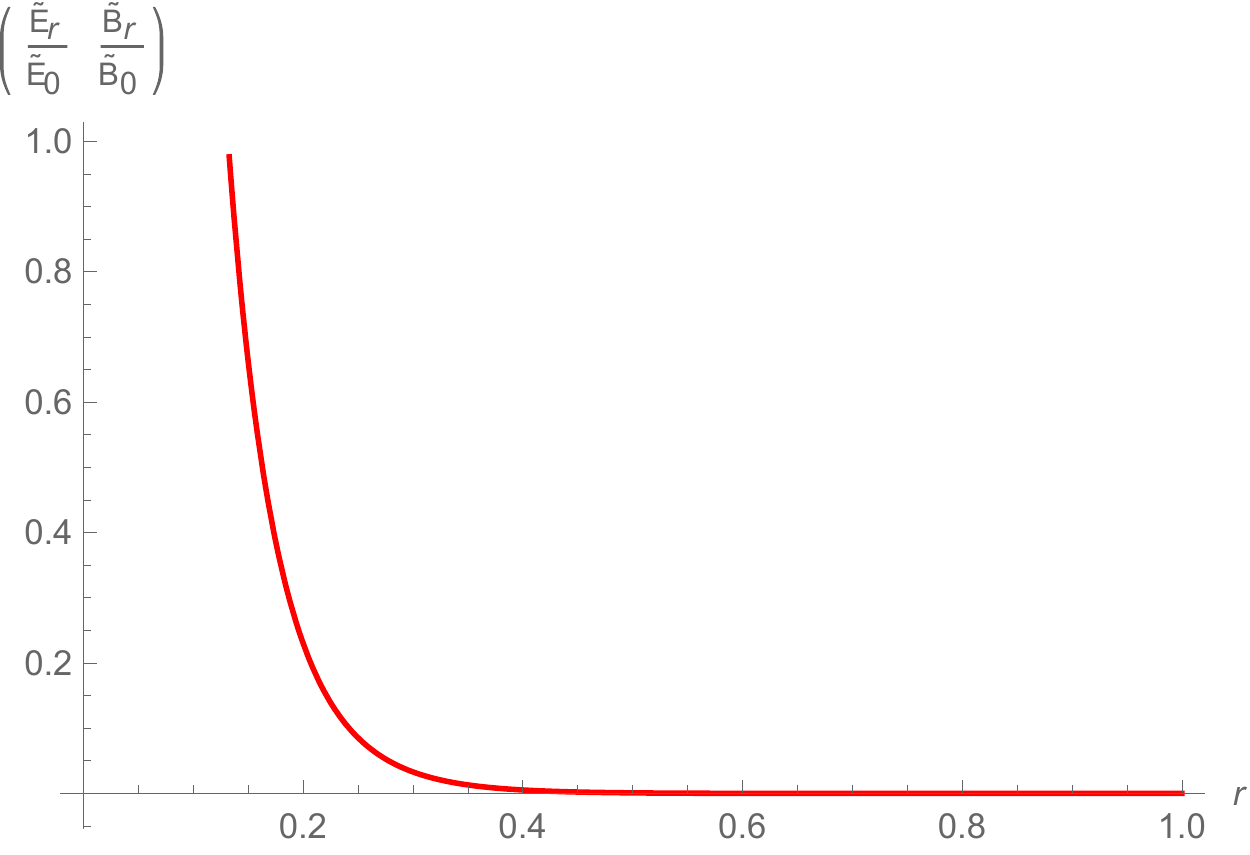}}
 \caption{The screened electric(magnetic) field $\tilde{E}_r(\tilde{B}_r)$ against $r$}
   \label{ef}
   \floatfoot{This graph holds for both electric and magnetic fields, thus, both fields are screened from the interior of the vacuum. It was sketched for $\lambda= 65\text{nm}$ below it, we have a mix or an intermediate states with higher electric or magnetic field strength. Above the estimated $\lambda$ we have a superconducting state because the vacuum is pure diamagnetic with no penetrating electric or magnetic fields.}
\end{figure}

\begin{figure}[H]
  \centering
 {\includegraphics[scale=0.6]{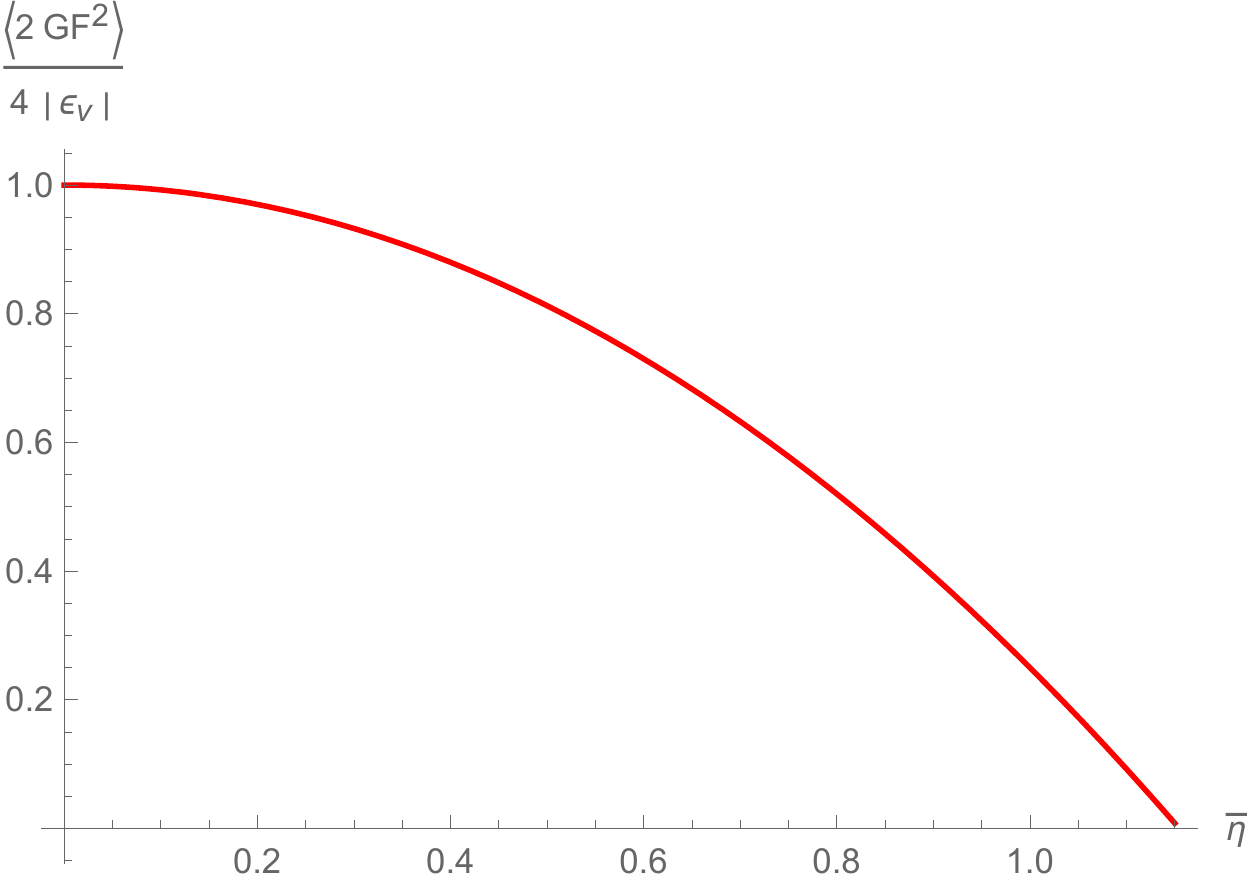}}
 \caption{The gluon condensate with the mean glueball field $\bar{\eta}$ }
   \label{vacuum}
   \floatfoot{The condensate has its maximum value at $\bar{\eta}=0$ and vanishes at $\bar{\eta}\simeq 1$. 
   }
\end{figure}



\begin{figure}[H]
  \centering
 {\includegraphics[scale=0.6]{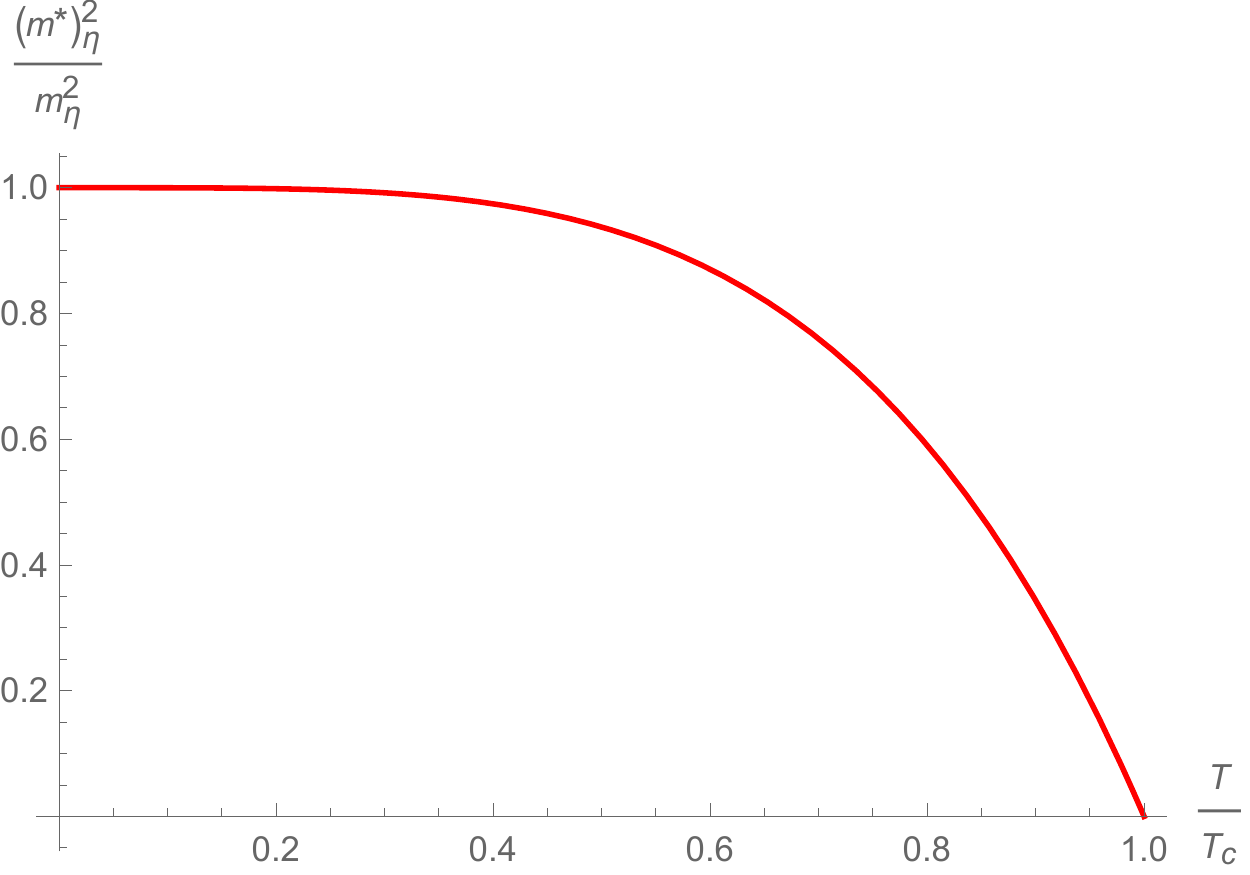}}
 \caption{Thermally fluctuating glueball mass with temperature}
   \label{gbm}
   \floatfoot{The glueball mass decreases sharply with increasing $T$ until it vanishes at $T=T_c$, where the glueballs melt into gluons.}
\end{figure}

\begin{figure}[H]
  \centering
 {\includegraphics[scale=0.6]{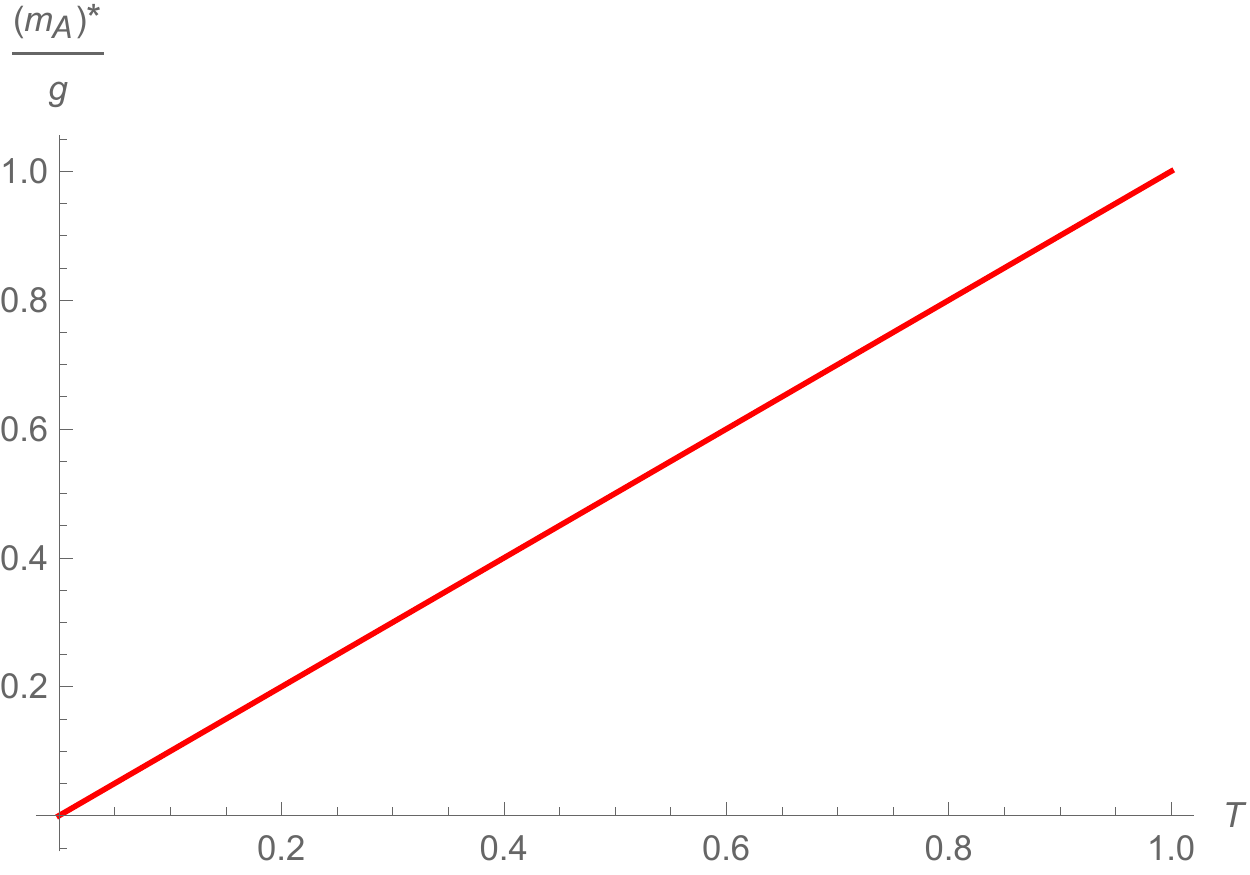}}
 \caption{Variation of the gluon mass $m_A^{*}(T)/g$ with $T$}
   \label{gm}
   \floatfoot{The gluon mass rather increases with temperature, because this phenomenon occurs at a temperature where the bound states of gluons or glueballs melt into gluons. 
   }
\end{figure}

\begin{figure}[H]
  \centering
 {\includegraphics[scale=0.6]{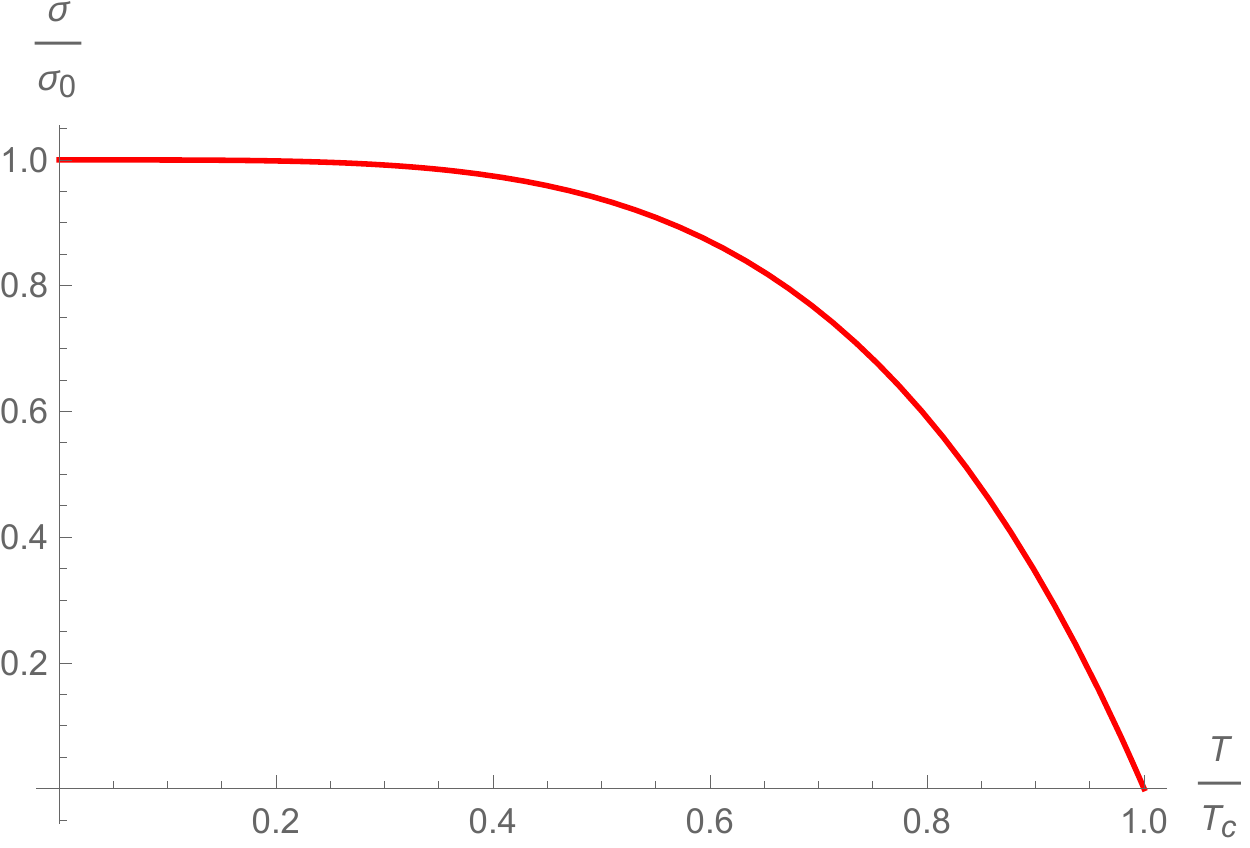}}
 \caption{The string tension $\sigma(T)/\sigma_0$ against temperature $T/T_c$}
   \label{st}
   \floatfoot{The string tension decreases sharply with $T$ and breaks or vanishes at $T=T_c$ representing hadronization.}
\end{figure}

\begin{figure}[H]
  \centering
 {\includegraphics[scale=0.6]{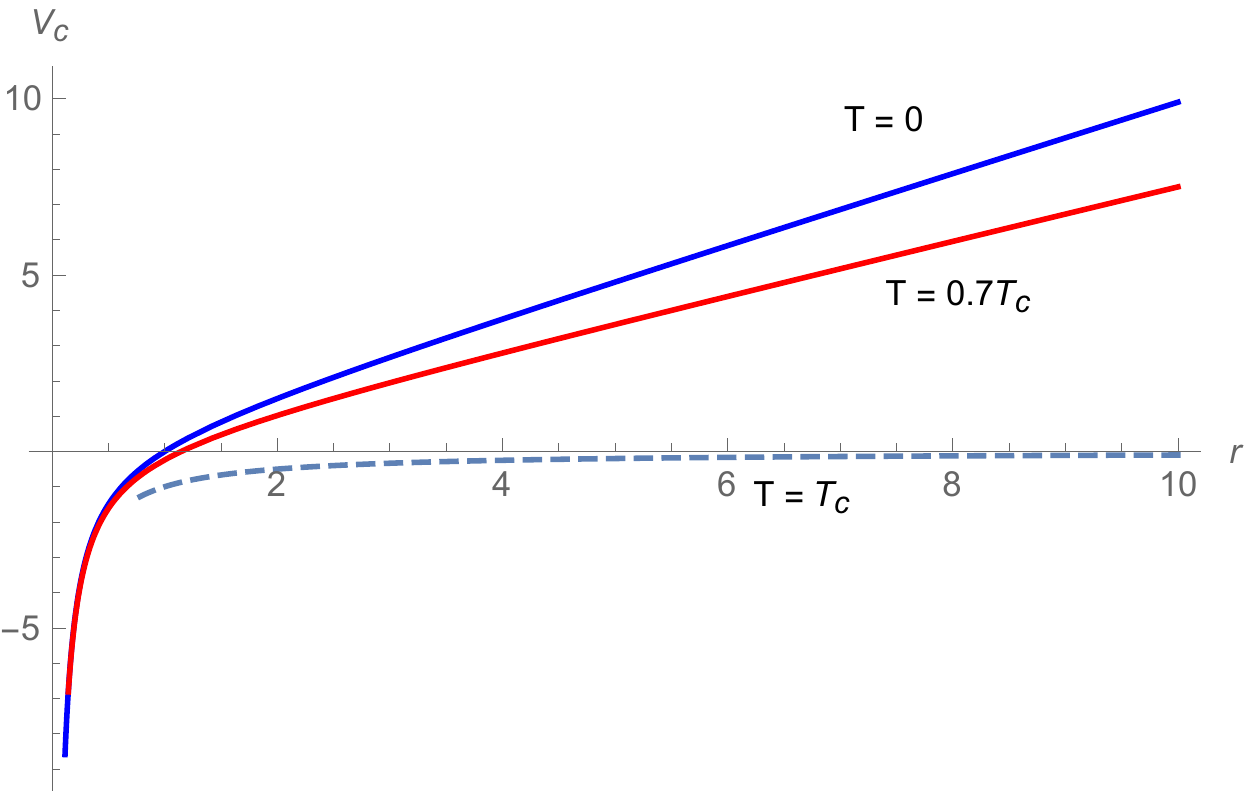}}
 \caption{Temperature fluctuating Cornell-like potential.}
   \label{cpt}
   \floatfoot{The gradient of the graph decreases with increasing $T$ and flattens at $T=T_c$ indicating deconfinement and hadronization phase.}
\end{figure}

\begin{figure}[H]
  \centering
 {\includegraphics[scale=0.6]{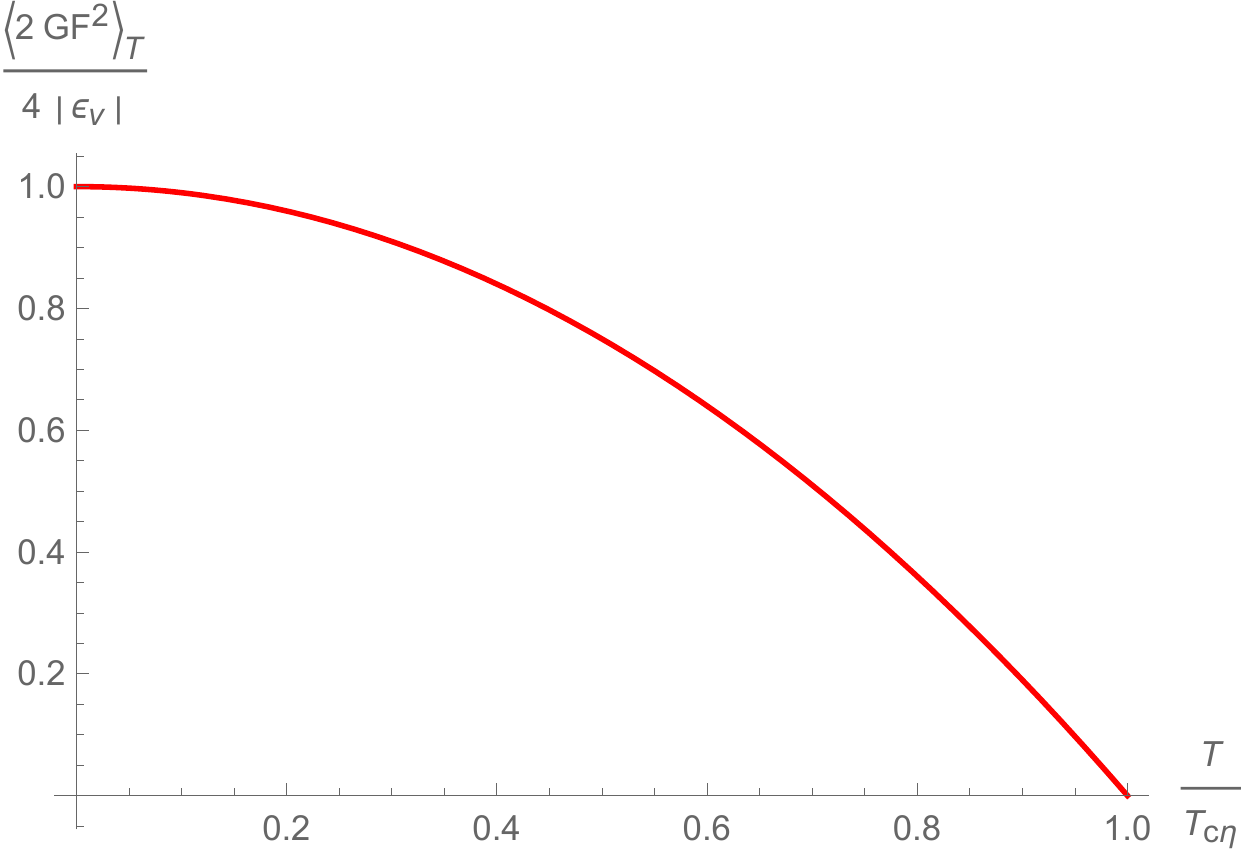}}
 \caption{The gluon condensate with varying temperature $T$}
   \label{vacuum2}
   \floatfoot{At $T=0$, $\langle 2G(\eta)F^{\mu\nu}F_{\mu\nu}\rangle_T/4|\varepsilon_v|=1$ representing maximum condensate corresponding to $\bar{\eta}=0$. Also, at $T=T_c$, $\langle 2G(\eta)F^{\mu\nu}F_{\mu\nu}\rangle_T/4|\varepsilon_v|=0$ representing minimum condensate corresponding to $\bar{\eta}\simeq 1$ \cite{Carter}.}
\end{figure}

\begin{figure}[H]
  \centering
  \subfloat[Left Panel]{\includegraphics[scale=0.6]{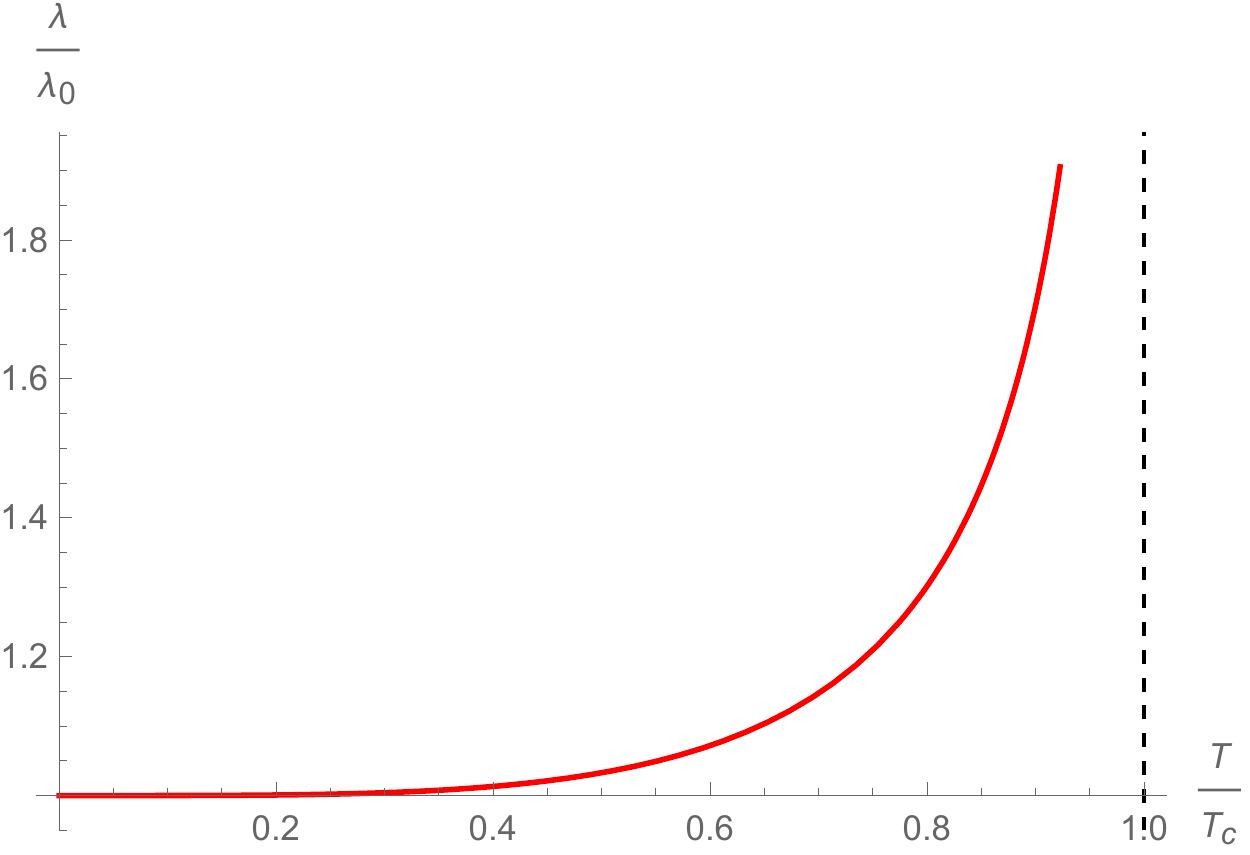}}
  \qquad
  \subfloat[Right Panel]{\includegraphics[scale=0.6]{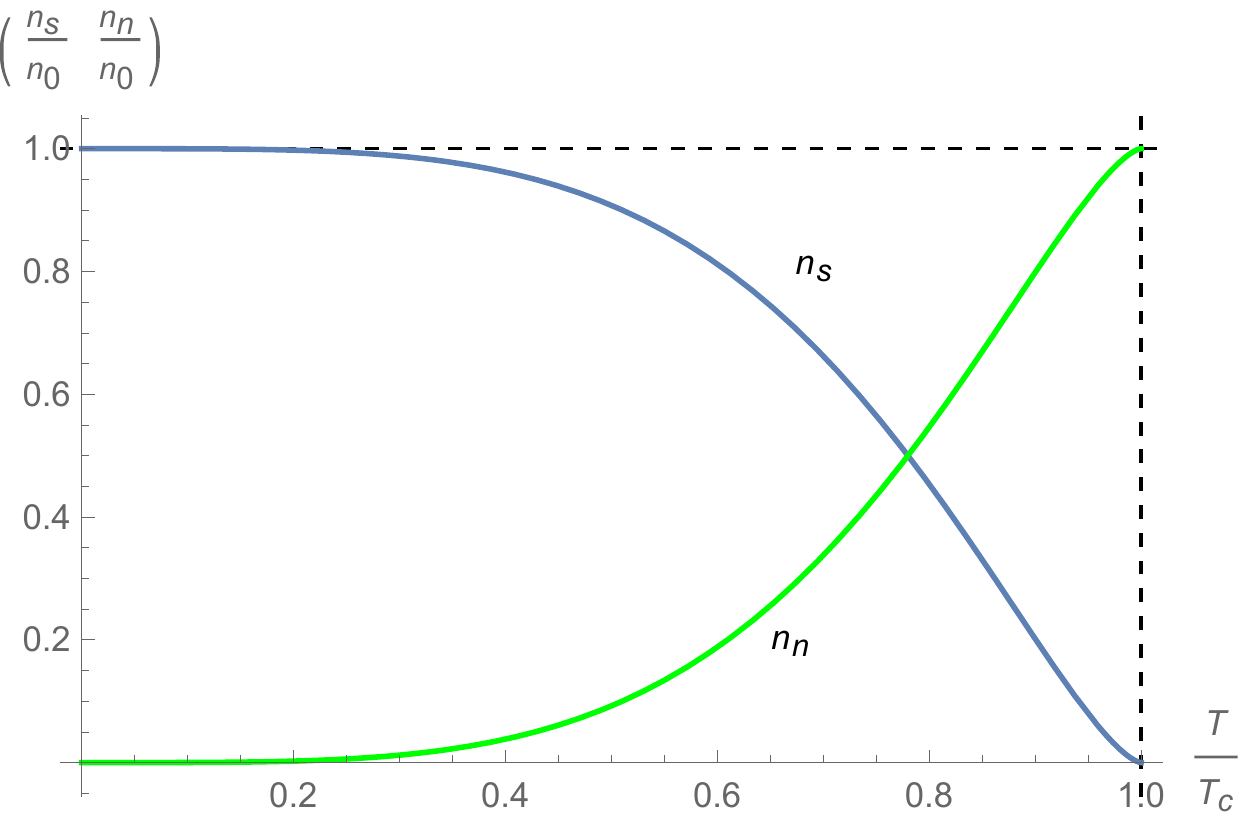}}
 \caption{Penetration depth $\lambda$ (left), superconducting $n_s$ and normal $n_n$ monopole densities with temperature (right)}
   \label{st2}
  \floatfoot{The penetration depth remains constant at $\lambda=\lambda_0$ representing pure diamagnetic vacuum before it starts rising steadily with $T$ and becomes infinite at $T=T_c$, where the fields are expected to penetrate the vacuum. Thus, we have superconducting state at low temperatures and at relatively high temperatures we have a mix or an intermediate states. On the other hand, the superconducting monopole density $n_s$ decreases with increasing $T$ while the normal monopole density $n_n$ increases with an increasing $T$.}
\end{figure}

\subsection{Conclusion}\label{conc}
{The process of high energy {annihilation of $\phi^*\phi$ to the production of $\phi^*\phi$ through the interaction, $\phi^*\phi\rightarrow \gamma\rightarrow {\phi}^*{\phi}$ during hadronization in high particle density region is briefly addressed}.} Thus, we focus on the discussion to cover the low energy regime where $\tilde{\text{U}}(1)$ symmetry is spontaneously broken through {\it Abelian Higgs mechanism} to give mass to the resulting glueballs. The scalar field in this case plays the role of Higgs field, which undergoes modification into glueballs with mass $m_\eta$. This leads to {\it color confinement} of glueballs. 
We explored the behaviour of the scalar field potential $V(|\phi|)$ and the glueball potential $V(\eta)$ with the decay constant $f_\alpha$ and the results presented in Figs.~\ref{sp} and~\ref{ep}. 
The model explains other confining (IR) properties such as gluon condensation, glueball mass, gluon mass, string tension and dual superconductivity through monopole condensation. The results for color confinement, screened electric (magnetic) field due to the monopoles and the gluon condensate are presented in Figs.~\ref{cp},~\ref{ef} and~\ref{vacuum} respectively. The glueball fields acquire their mass through SSB and their mass remain the most relevant parameter throughout the confinement properties. The magnitude of the glueball mass $m_\eta$ is precisely the same as the observed lightest scalar glueball mass $m_\eta=1.73\,\text{GeV}$. This mass is responsible for the string tension $\sigma$ that keeps the particles in a confined state and appears in the monopole condensate as well. It also appears in the gluon condensate $\langle 2G(\eta)F^{\mu\nu}F_{\mu\nu}\rangle$, capable of hadronizing to form light pions in the low energy regime. {In this paper two forms of hadronization were discussed, the first one takes place at the high energy regime where the individual quarks and gluons recombine to form hadrons/pions. Secondly, at the low energy regime where the quark and an antiquark pairs/valence gluons that form pions/glueballs get separated due to long separation distances and the string tension that hold them together breaks leading to hadronization.} We also investigated the effect of temperature on the effective scalar glueball mass $m_\eta^{*2}(T)$, the gluon mass $m_A^{*}(T)$, string tension $\sigma(T)$, confining potential $V_c(r,T)$, gluon condensate $\langle 2G(\eta)F^{\mu\nu}F_{\mu\nu}\rangle_T$, the penetration depth $\lambda(T)$, the superconducting monopole densities $n_s(T)$, the normal monopole density $n_n(T)$ and their results presented in Figs.~\ref{gbm}, ~\ref{gm},~\ref{st},~\ref{cpt} ~\ref{vacuum2} and ~\ref{st2} respectively. We calculated the QCD $\beta$-function and the strong `running' coupling $\alpha_s$ through renormalization group theory to enhance the discussions on gluon mass generation. 
Additionally, the glueball mass and the gluon masses were calculated and the outcome compared with lattice simulation result, analytical study or phenomenological analyses to ascertain their reasonability. 

Finally, the model produces the correct behaviour of confining potential at $T=0$, where the potential has linear growth with $r$, Eq.(\ref{12a}), consistent with Cornell potential model. The potential keeps growing linearly for $T\,<\,T_c$ with thermal deconfinement phase at $T\,\geqslant\,T_c$. However, most of the results on QCD lattice calculations point to a temperature correction of order $-T^2$ to the string tension \cite{Kaczmarek,Pisarski,Forcrand} meanwhile, the model proposes correction in order of $-T^4$ in Eq.(\ref{ana1}). This can be regularized at $T\simeq T_c$ to obtain the correct order $-T^2$. Indeed, we did not find such regularization necessary because there are some proposed phenomenological models that suggest $-T^4$ correction \cite{Boschi-Filho} to the string tension as well. But we deem it necessary to provide some explanations based on the model. In the previous paper Ref.\cite{A-B} we obtained a correction of order $-T^2$ as elaborated in Eq.(\ref{stringt}). We observed that in Ref.\cite{A-B} the temperature correction to the string tension came from the scalar field $\phi$ as elaborated in the latter part of Sec.~\ref{gc}. On the other hand, the temperature correction in the gauge field leads to a correction in the order of $-T^4$ as discussed in Sec.~\ref{em} and elaborated in Sec.~\ref{ana} in Eqs.(\ref{ana1}) and (\ref{ana2}). 

\acknowledgments

We would like to thank CNPq, CAPES and CNPq/PRONEX/FAPESQ-PB (Grant no. 165/2018), for partial financial support. FAB also acknowledges support from CNPq (Grant no. 312104/2018-9).

\end{document}